\documentclass[aps,pre,twocolumn,superscriptaddress,nofootinbib]{revtex4-2}

\usepackage{soul}

\usepackage{amsthm} 
\usepackage{amsmath} 
\usepackage{amssymb}
\usepackage{afterpage} 
\usepackage{graphicx}
\usepackage{xcolor}
\usepackage{xr-hyper}
\usepackage[colorlinks=false]{hyperref}
\usepackage[capitalise]{cleveref}

\definecolor{RED}{rgb}{1,0,0}
\definecolor{BLACK}{rgb}{0,0,0}

\newcommand{\rr}[1]{\textcolor{black}{#1}}

\begin{document}

\title{Scientific productivity as a random walk} 

\author{Sam Zhang}
\email{sam.zhang@uvm.edu}
\affiliation{Department of Applied Mathematics, University of Colorado, Boulder CO 80309, USA}
\affiliation{Department of Mathematics and Statistics, University of Vermont, Burlington VT 05405, USA}
\affiliation{Santa Fe Institute, Santa Fe, NM 87501, USA}

\author{Nicholas LaBerge}
\affiliation{Department of Computer Science, University of Colorado, Boulder CO 80309, USA}

\author{Samuel F. Way}
\affiliation{Department of Computer Science, University of Colorado, Boulder CO 80309, USA}

\author{Daniel B. Larremore}
\affiliation{Department of Computer Science, University of Colorado, Boulder CO 80309, USA}
\affiliation{BioFrontiers Institute, University of Colorado, Boulder CO 80309, USA}
\affiliation{Santa Fe Institute, Santa Fe, NM 87501, USA}

\author{Aaron Clauset}
\email{aaron.clauset@colorado.edu}
\affiliation{Department of Computer Science, University of Colorado, Boulder CO 80309, USA}
\affiliation{BioFrontiers Institute, University of Colorado, Boulder CO 80309, USA}
\affiliation{Santa Fe Institute, Santa Fe, NM 87501, USA}

\begin{abstract} 
The expectation that scientific productivity follows regular patterns over a career underpins many scholarly evaluations. 
However, recent studies of individual productivity patterns reveal a puzzle: the average number of papers published per year robustly follows the ``canonical trajectory'' of a rapid rise followed by a gradual decline, yet only about 20\% of individual productivity trajectories follow this pattern. 
We resolve this puzzle by modeling scientific productivity as a random walk, showing that the canonical pattern can be explained as a decrease in the variance in changes to productivity in the early-to-mid career. 
By empirically characterizing the variable structure of 2,085 productivity trajectories of computer science faculty at 205 PhD-granting institutions, spanning 29,119 publications over 1980--2016, we (i) discover remarkably simple patterns in both early-career and year-to-year changes to productivity, and (ii) show that a random walk model of productivity both reproduces the canonical trajectory in the average productivity and captures much of the diversity of individual-level
trajectories\rr{, including the lognormal distribution of cumulative productivity observed by William Shockley in 1957}.
\rr{We confirm that these results generalize across fields by fitting our model to a separate panel of 22,952 faculty across 12 fields from 2011 to 2023.}
These results highlight the importance of \rr{variance} in shaping individual scientific productivity, opening up new avenues for characterizing how systemic incentives and opportunities can be directed for aggregate effect. 
\end{abstract}

\maketitle

\section{Introduction}

Scientific productivity, which is typically measured by the number of papers that a scholar publishes, underpins many evaluative processes over the course of an academic career, including hiring decisions, tenure and promotions, grant funding, and scientific prizes~\cite{cole1967scientific, stephan2012economics}. 
Due to its broad importance, scientific productivity has been studied from a variety of angles, such as 
productivity over time, averaging over scholars~\cite{lehman1953age, levin1989age, gingras2008effects}; 
productivity over scholars, averaging over time~\cite{lotka1926frequency, price1963little}; 
and extremal statistics of the most productive or impactful papers or years within careers~\cite{sinatra2016quantifying, liu2018hot}. 
While useful, these approaches leave unanswered key questions about scientific careers that depend on knowledge about the full distribution of scholarship.

For example, a substantial literature\rr{, spanning many decades, fields, and countries}, documents a ``canonical trajectory'' in scientific productivity over a career. 
The canonical trajectory is when a researcher’s productivity tends to rise rapidly to a peak in the early career followed by a gradual decline, a pattern which is robustly captured when many scientists' trajectories are averaged~\cite{dennis1956age, diamond1986life, cole1979age, gingras2008effects}\rr{, in dozens of countries~\cite{lehman1953age}, including recent evidence from countries as diverse as Norway~\cite{rorstad2015publication} and
Spain~\cite{costas2010bibliometric}.}
However, recent work has revealed that this canonical trajectory is not representative of most individual scientists, who instead exhibit a rich diversity of productivity trajectories~\cite{way2017misleading}, even as their average productivity reliably follows the canonical trajectory.

The discovery that the canonical trajectory is a misleading description of individual productivity patterns presents a puzzle: what mechanisms lead to both dramatic variability in individual productivity trajectories and simultaneously the canonical pattern in aggregate? 
Past explanations of a canonical pattern at the individual level have invoked ideas ranging from cognitive mechanisms~\cite{horner1986relation} to psychological development~\cite{mumford1984age} and economic mechanisms~\cite{becker2009human, diamond1986life}. 
\rr{A second category of explanation focused on tenure, whereby faculty become less productive after tenure due to changing research incentives~\cite{holley1977tenure, brogaard2018economists, tripodi2025tenure} and a marked increase in administrative duties~\cite{myers2023newfacts}.}
\rr{A third set of explanations focus on the scientific reward mechanisms, in which scholars tend to become more stratified over the course of a career~\cite{cole1974social, reskin1977scientific, cole1979age, checchi2021incentives}}. 
However, these ideas \rr{generally} ignore the broad heterogeneities across scientists and institutions, and do not readily explain the empirical diversity of faculty productivity patterns~\cite{way2017misleading}. 
As a result, little is known about mechanisms that generate realistic individual productivity trajectories. 

Here, we propose and investigate a parsimonious explanation which links several simple observations by modeling scientific productivity as a discrete-time Markov chain, which we refer to as a random walk. 
First, individual faculty productivity fluctuates from year to year due to individually contingent factors and events, including the beginning of a new collaboration~\cite{li2022untangling, lee2005impact}, an experiment that fails~\cite{wang2019setback}, parenthood~\cite{morgan2021unequal}, or changing institutions~\cite{long1978productivity,zhang2022labor}. 
While individually unpredictable, these fluctuations combine to form recognizable statistical patterns in the aggregate.
Second, these factors change over a career, such that the variability of fluctuations also changes across different career stages, with higher productivity fluctuations in the early career than in the later career. 
In fact, we will show that a random walk with a change in variance is sufficient to produce both the canonical trajectory and much of the observed variability around it. 
\rr{This change in variance explanation builds on past work that highlights the relationship between institutional forces and systemic incentives on the one hand and global patterns of productivity on the other~\cite{cole1979age, reskin1977scientific}; 
on work that emphasizes the central role of randomness and luck in scientific careers~\cite{clauset2017data}, e.g., the unpredictability of when faculty tend to publish their most highly cited papers~\cite{sinatra2016quantifying, liu2018hot}; and on work that points toward the importance of changes of variance within faculty careers~\cite{hamermesh2023older} and other trajectories~\cite{gould1988trends}.}

We formalize this explanation as a probabilistic generative model that can simulate the evolution of individual faculty productivities, which we validate against \rr{two data sets: comprehensive empirical data on the productivities of 2,085 computer scientists at PhD-granting universities in the US and Canada; as well as the first decade of 22,952 tenure-track faculty careers across 12 fields}. 
We produce \rr{four} models---\rr{two baseline models without a random walk, a simplified random walk model, and a full random walk model}.

\rr{The baseline models show that two intuitive and tempting approaches to modeling faculty trajectories---an exponentially distributed productivity around a time-varying mean, or productivity as arising through a basket of projects each with Poisson maturation times---both fall short of key characteristics of the empirical data.
On the other hand,} a simplified \rr{random walk} model shows that a change of variance in faculty careers is sufficient to produce the canonical trajectory while preserving individual variability. 
It crystallizes a set of sufficient conditions for producing canonical patterns, and allows us to explore the space of possible average trajectories. 
The full model shows that modeling productivity as a random walk captures many of the details of both individual productivities, and aggregate patterns like the canonical trajectory \rr{and a cumulative lognormal productivity distribution across individuals}~\cite{shockley1957statistics}, while simultaneously revealing noteworthy certain non-Markovian patterns in real faculty productivity.

The full model fits two sets of parameters: the change points between career stages, which parameterizes the change of structural influences across a scientific career, and the parameters describing the distribution of productivity fluctuations within each career stage, which parameterizes the role of contingency and luck. 
Together, these assumptions model an individual researcher's productivity over time as a truncated random walk that cannot become negative, where individual step sizes are drawn from a distribution whose parameters depend on the individual's career stage.

We first show that the simplified model is sufficient for generating a diverse range of trajectories that reproduce the canonical trajectory in aggregate. 
We then fit the full model to the empirical data on computer scientists and obtain estimates of the model's change points, which represent the timings of major career transitions for faculty researchers, and the parameters for the random walk within each career stage. 
We directly validate the timing of the inferred career change points by comparing them to the typical timing of faculty promotions for this population of researchers. 
We then check the fitted model by generating an ensemble of simulated productivity trajectories, which we contrast with the empirical trajectories across a variety of statistical measures. 
The full model successfully explains a substantial portion of the variability of individual careers as well as the canonical trajectory pattern, while also revealing important discrepancies between the model and the data that indicate higher-order mechanisms and other contingent forces that shape scientific productivity. 
\rr{Lastly, we validate the generalizability of the approach by fitting the full model to all 12 fields.}

\begin{figure*}[htb!] 
\centering 
\includegraphics[width=1\linewidth]{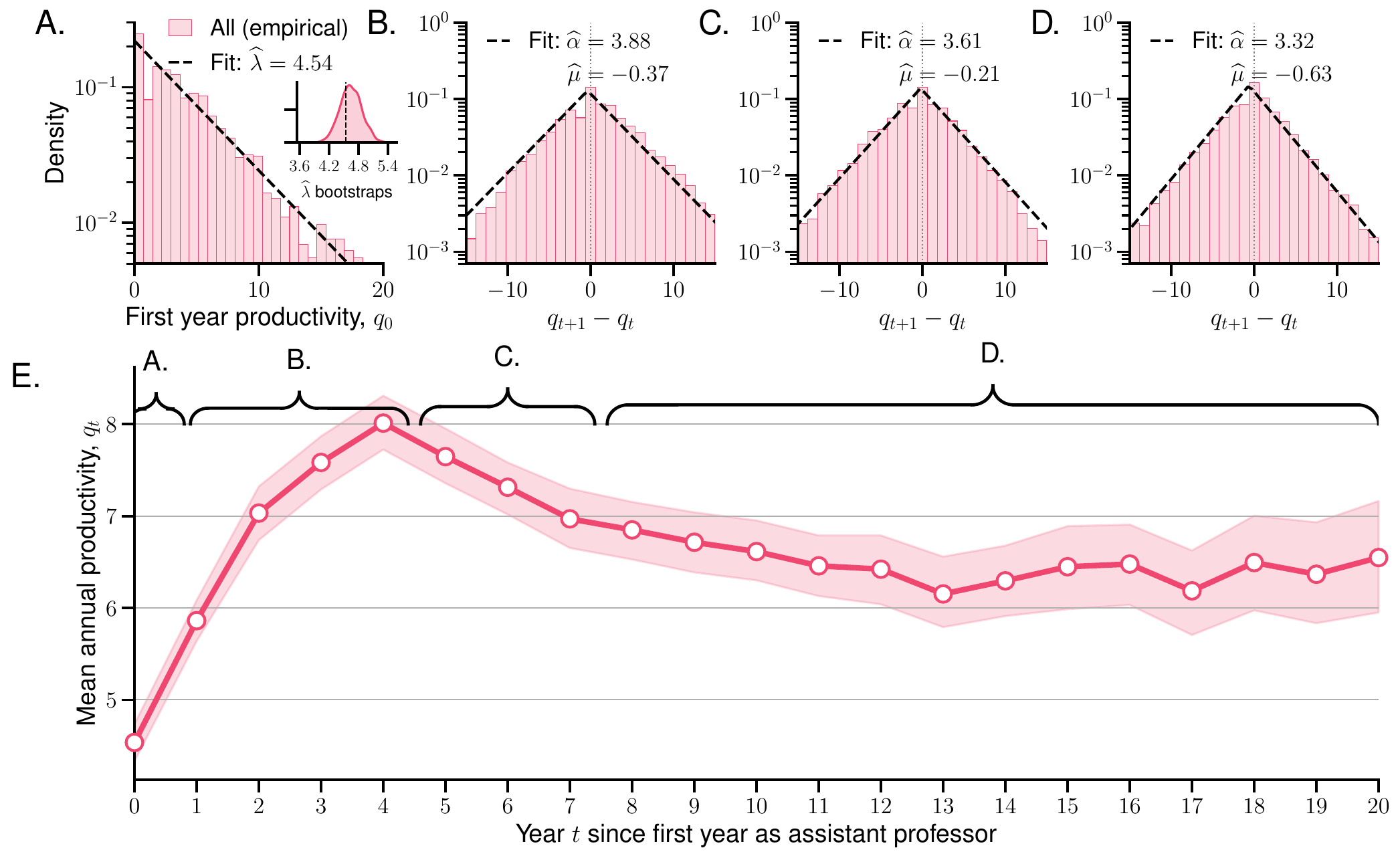} 
\caption{\textbf{Empirical productivity data}. 
(A) An exponential distribution (dashed black line) accurately fits the empirical first-year productivity (pink histogram). 
The inset displays the estimated rate parameter against the density of estimated rates in 1,000 bootstrap replicas. 
(B-D) The empirical distributions of productivity changes (pink histograms) are semi-log plots, for ranges of career age, along with fitted Laplace distributions (dashed black line). 
(E) The average productivity for the same set of researchers, showing the ``canonical trajectory'' of a rapid rise followed by a gradual decline or leveling off, depicted as means of time-adjusted productivity for each career age and 95\% bootstrap confidence intervals. Brackets indicate the range of career ages that were grouped together for the density plots: (A) productivity in year zero, and then changes of productivity in (B) years 1--4, (C) years 5--7, and (D) years 8--20. 
} 
\label{fig:criteria} 
\end{figure*}

\section{Data} 

We combine three comprehensive datasets to perform our analysis. 
First, we use a hand-curated census of all tenured or tenure-track faculty employed at all 205 US and Canadian computer science departments documented in the Computing Research Association (CRA)'s Forsythe List of PhD-granting departments in computing-related disciplines~\cite{cra2012forsythe} in the academic year 2011--2012. 
This dataset \rr{contains the career trajectories of} 5,032 faculty \rr{up to the year 2011}, whose PhD degree institutions and employment histories were manually gathered from public materials such as CVs and academic websites. 

Second, we use the November 2016 snapshot of the Digital Bibliography and Library Project (DBLP, ~\cite{dblp2016november}), a large-scale bibliographic dataset for journals and conference proceedings relevant to computing research, although with limited coverage of interdisciplinary computing. 
The employment data is joined with the DBLP both algorithmically and manually, excluding preprints on the arXiv. 
By using publication data linked to definitive employment records, rather than inferring the start of careers from publications, as is common in the bibliometrics literature~\cite{wang2017scientific,kwiek2022academic, gyHorffy2022evaluating}, we are able to isolate and analyze the dynamics of scholarly productivity under a relatively consistent and stable set of influences and incentives around productivity. 

To account for DBLP's degraded coverage of publication records further back in time and non-stationarity in average productivities over time, we use the linear scaling developed by Way et al.~\cite{way2017misleading} that adjusts the average productivity in DBLP to match the average productivity estimated from a random sample of CVs from the same population of researchers. 
This adjustment allows us to include researchers from different career stages into a single analysis, and to compare faculty at a similar career stage across cohorts. 
This adjustment results in a real-valued non-negative number for each faculty in each year $t$ that we will denote as the \emph{adjusted productivity} $q_t$.
We denote the change in adjusted productivities as $\delta_t = q_{t+1} - q_t$.

We focus our analysis on the most productive years of a career, and where the population pattern of the canonical trajectory is strongest, by analyzing years 0--20 of the careers for all faculty who received their PhD on or after 1980. 
We refer to the number of years since the start of a professor's first assistant professorship as their \emph{career age}, with their first year as career age $0$. 

To be included in our analysis, we require that faculty publish three or more papers indexed by DBLP before career age $5$. 
These inclusion criteria result in a dataset of 2,085 faculty across $204$ departments, and $128,816$ author-publication pairs. 
For a subset of our analyses, we select faculty whose careers span the full 21 years, which yields $510$ careers. 
We designate these careers the \emph{full trajectories}.

\rr{Third, for the cross-field analysis, we use faculty employment panel data from 2011 to 2023 supplied by the Academic Analytics Research Center (AARC), linked to their publication records in Scopus.
    Relying on the completeness of the AARC data, we infer the start date of employment for faculty with the ``Assistant Professor'' job title based on the first year they appear in the dataset.
    Among individuals with an inferred start year who had at least one publication linked in Scopus, we end with 102,473 years of data across 22,952 individuals, ranging from 1,204 scholars in sociology to 2,669 in psychology (see Supporting Information for more details on this data).}

\begin{figure*}[htb!] 
\centering 
\includegraphics[width=1\linewidth]{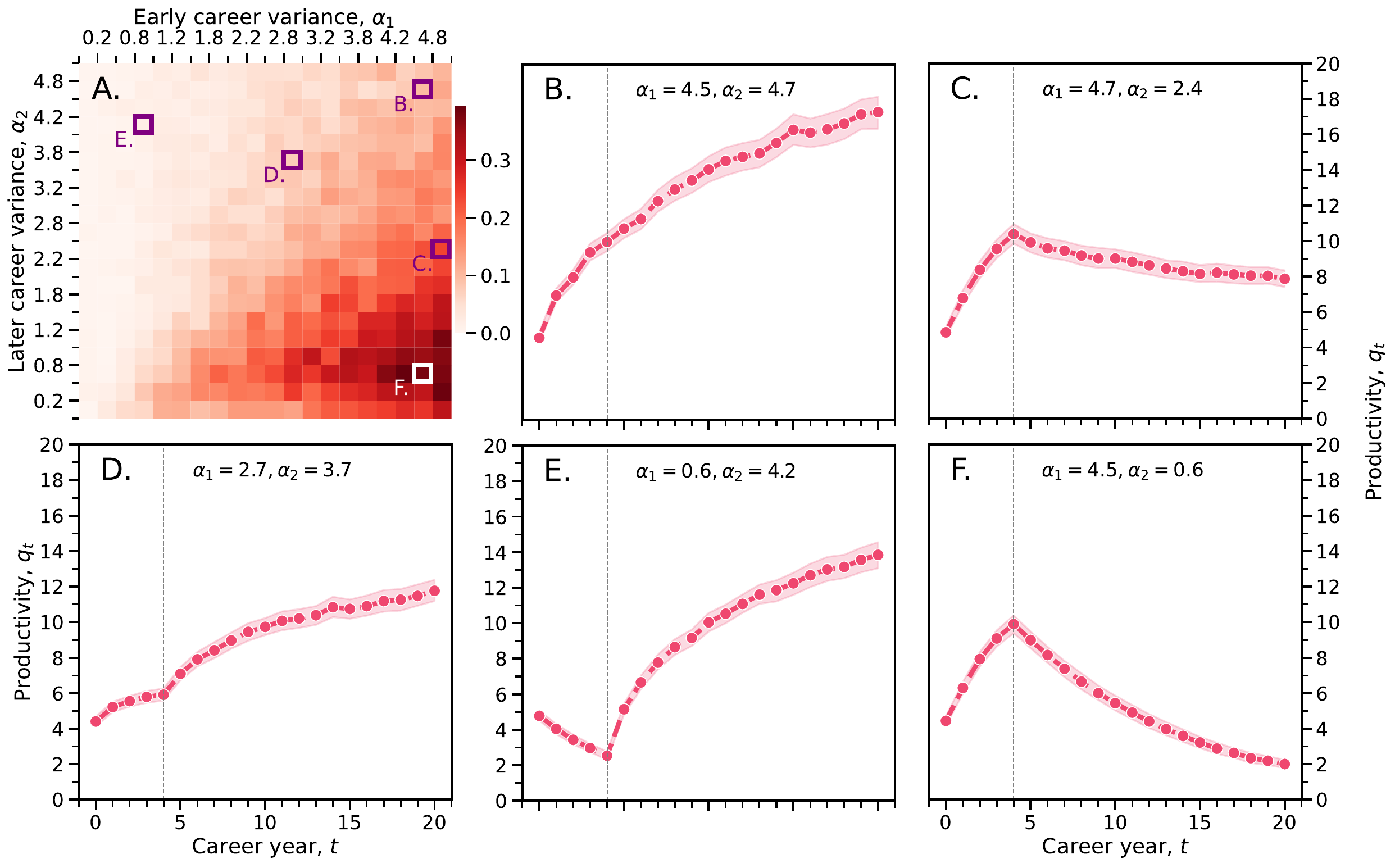} 
\caption{Reproducing canonical trajectories with a simplified model. 
(A) Simulating $N=400$ trajectories for each pair of $\alpha_1$ and $\alpha_2$ with $\mu=-1$ fixed, we display the fraction of those trajectories that are canonical. Some regions of the parameter space generate non-canonical trajectories (B, D, E), while others generate more canonical trajectories on average (C, F). Shaded intervals denote pointwise 95\% confidence intervals for $N=1000$ simulations at those parameters. 
} 
\label{fig:simplified_canonical} 
\end{figure*} 

\section{Results} 

\subsection{Distribution of productivity changes} 
To study faculty careers from a perspective beyond average or extreme values, we characterize the stochasticity and variation within and across individuals by examining how productivity varies at the start of a career, and how it evolves empirically over time. 
We examine the distribution of first-year productivity $q_0$, and the distributions of changes in productivity $\delta_t = q_{t+1} - q_t$, and find surprising statistical regularity in both distributions: first-year productivity closely follows an exponential distribution (\cref{fig:criteria}A), and the productivity changes follow a Laplace distribution regardless of career stage~(\cref{fig:criteria}B-D). 
The simple form of these empirical distributions is provocative, and suggests that the variability of initial productivity $q_0$ and subsequent changes to productivity $\delta_t$ may reflect relatively simple underlying stochastic processes.

Fitting exponential and Laplace distributions to the data, we notice that the estimated variances decrease from $\widehat{\alpha}=3.88$ (95\% CI: [3.78, 3.97]; all CIs are individual-level block bootstraps with 10,000 bootstrap replicas) to $\widehat{\alpha}=3.64$ (95\% CI: [3.54, 3.77]) and $\widehat{\alpha}=3.32$ (95\% CI: [3.19, 3.39]) over the course of a career~(\cref{fig:criteria}). 
On the other hand, the location parameters exhibit much more inferential uncertainty as well as the lack of any clear pattern, where between career years 1--4 and 5--7, the mode increases from $\widehat{\mu}=-0.37$ (95\% CI: [-0.48, 0.35]) to $\widehat{\mu}=-0.21$ (95\% CI: [-1.77, 0.32]), despite a change in the average trajectory from increasing to decreasing. 
This pattern suggests that the variance, rather than the location, of these distributions, plays the key role in shaping the appearance of the canonical trajectory. 
The fact that across all career stages $\hat{\mu} < 0$ is intriguing, as it suggests a downward pressure on productivity over time, i.e., the mode of next year's productivity will be slightly lower than this year's.

\subsection{\rr{Evaluating baseline models}}
\rr{We evaluate two baseline models with intuitive appeal to confirm that they fall short in substantial ways that require us to develop our more sophisticated dynamical model. The first baseline directly models the canonical trajectory using a piece-wise linear model~\cite{way2017misleading}, and then productivity is independently drawn from exponential distributions (see Supporting Information).
Compared to empirical data, this baseline model heavily underestimates the probability of most productive years occurring at the beginning of a career, and overestimates its probability in the later career~(\cref{fig:minimal_model1}).}

\rr{The second baseline assumes that faculty initiate $K$ projects at the same time, and each project matures based on a Poisson distribution with rate $\lambda$.
    When a project matures, it yields a paper, and the project resets.
    Simply through the initial synchronization of several projects starting at the same time, and their subsequent decoherence, we observe canonical trajectories for certain sets of parameters $K$ and $\lambda$~(\cref{fig:minimal_model2}A).
However, the increments of productivity are not Laplace-distributed~(\cref{fig:minimal_model2}B,D), and the canonical trajectory does not consistly appear across commonsense regions of the parameter space~(\cref{fig:minimal_model2}C).} 

\subsection{Modeling the canonical trajectory} 
Given the statistical regularity of the $q_0$ and $\delta_t$ distributions, we test whether changes in variance could drive the shape of the canonical trajectory by building a simple model.
To do so, we build on the literature suggesting simple two-stage careers---that faculty productivity experiences a qualitative transformation around tenure, with rapid rise before and gradual decline after---to construct a simplified model with separate variance parameters for either stage~\cite{lehman1953age, dennis1956age, diamond1986life, cole1979age, gingras2008effects}. 

Our simplified random walk model of the productivity of a faculty career is a discrete-time Markov chain with two free parameters: the variance in the early career $\alpha_1$ (before year 5), and the variance in the later career $\alpha_2$ (after year 5). 
Following our empirical observation that the mode is typically negative, we fix the mode of the distribution at $\mu = -1$. 
By simulating career trajectories at each pair of possible variances $(\alpha_1, \alpha_2)$, we examine whether there exist necessary criteria on the variances of faculty productivity for producing canonical trajectories at the individual level. 

Across the parameter space, we find that high variance in the early career paired with low variance in the later career \mbox{$\alpha_2 < \alpha_1$}, reliably produces a canonical trajectory at the individual level~(\cref{fig:simplified_canonical}C,F), while other choices of variances typically do not~(\cref{fig:simplified_canonical}B,D,E). 
In contrast, low variance in the early career followed by a higher variance later \mbox{$\alpha_1 < \alpha_2$} tends to produce an aggregate trajectory with a ``bounce'', in which the average productivity falls to an early nadir, and then gradually rises over time. 
When the variances are equal or nearly so, the average productivity instead tends to rise to a level that is proportional to the variance's magnitude. 
Finally, regardless of the parameterization, most individual trajectories do not follow the corresponding aggregate trajectory, and instead individual trajectories exhibit the broad diversity of shapes observed in empirical data~\cite{way2017misleading}. 

The appearance of the canonical trajectory when \mbox{$\alpha_1 > \alpha_2$} occurs for a straightforward mathematical reason: because the random walk tends to drift toward zero ($\mu=-1$), but productivity cannot be negative ($q_t \ge 0$), the Markov chain's expected value will tend to relax onto a value that is roughly proportional to the variance. 
(We derive this behavior analytically in the Supporting Information.) 
The canonical pattern appears because initial productivity $q_0$ is close to zero, causing the average productivity to rise initially. 
But, because $\alpha_1 > \alpha_2$, the Markov chain overshoots the expected productivity of the later career period, and at the beginning of that period, when the variance shifts to its lower value, the expected productivity then gradually falls. 
Hence, the canonical pattern can be explained as a natural consequence of a reduction in the variance of annual productivity over a career. 

\begin{figure*}[htb!] 
\centering 
\includegraphics[width=1\linewidth]{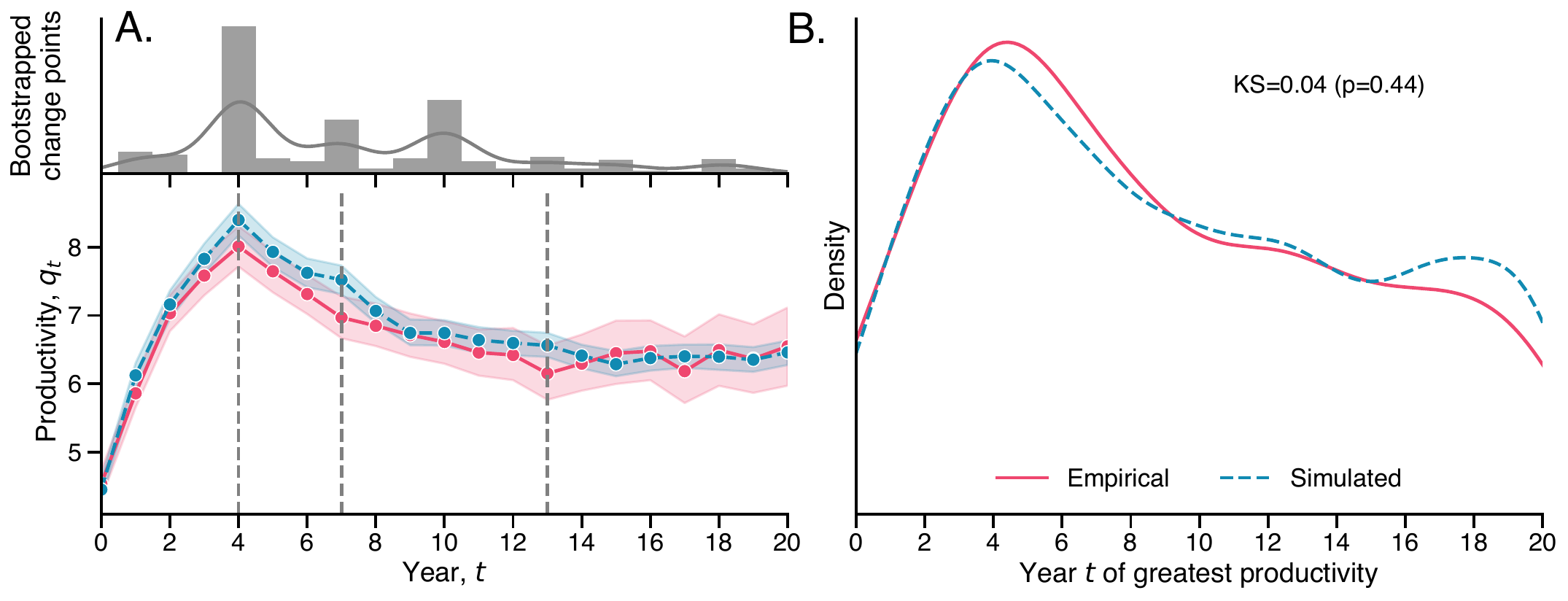} 
\caption{\textbf{Fitting the empirical data} 
(A) Average productivity by career year for real and simulated trajectories, where shaded ribbons denote 95\% confidence intervals. 
Dashed gray lines denote estimated career change points (at years 4, 7, and 13). 
Above, the bootstrap distribution of change points across $1000$ bootstrap iterations, where bootstrap is conducted at the individual level. 
(B) Distribution of the years with greatest productivity among the full empirical and simulated trajectories. 
Distributions are similar across the entire career ($\text{KS}=0.04$; $p=0.44$). 
} 
\label{fig:fit} 
\end{figure*} 

\subsection{Modeling empirical productivity trajectories} 

While the simple model confirms that a change in variance is sufficient to produce a canonical trajectory in a two-stage career, real productivity trajectories may exhibit more than two stages. 
We therefore introduce a full model that decides on the number of career stages from the data, as well as the years spanned by each stage. 
To prevent overfitting to the data by adding overly many career stages, we regularize this model by fitting a productivity-dependent mode that allows greater shrinkage from high productivity values (see Supporting Information).

In this model, initial productivity is drawn from an exponential distribution with rate $\widehat{\lambda}_0$, and we estimate the number and location of breakpoints between career stages. 
In each career stage $i$, we further fit both scale $\widehat{\alpha_i}$ and location slope $\widehat{\beta_i}$ for the Laplace distribution governing the change in productivity, such that the conditional probability of observing a change in productivity $\delta$ following a year with productivity $x$ is given by:

\begin{equation*}
f( x, \delta ) =  \mathbb{I}\lbrace \delta > -x \rbrace ( 2 - e^{ - \beta_i x/\alpha_i })^{-1} ( 1/\alpha_i ) e^{-|\delta - \beta_i x|/\alpha_i}
\end{equation*}

These parameters can be accurately and efficiently estimated from data, and we confirm this fact by recovering known parameters, including career breakpoints, from simulated data (see Supporting Information). 

\textbf{Fitted parameters.} 
Despite the full model's increased complexity relative to the simplified model, its estimated parameters remain fully interpretable. 
The estimated career stages denote regimes with similar productivity dynamics, meaning a relatively stable set of factors, both systematic and contingent, that influence a scientist's productivity. 

After fitting the full model to the set of 2,085 productivity time series in our data, we perform an initial check of the model's fit by examining the estimated parameters. 
The maximum likelihood fit yields four career stages: years 0--4, 5--7, 8--13, and 14--20~(\cref{fig:fit}A). 
These inferred career stages align well with common transitions that correspond to promotions in faculty careers, such as tenure evaluation which typically occurs in career years 5--7, and promotion to full professor, which often occurs about 12--15 years into a faculty career~\cite{spoon2023retention}. 
We note that the inferred change points varied across bootstrap replicas, with no set of maximum likelihood change points occurring in over 13\% of replicates. 
The change points our procedure infers from the empirical data (4, 7, and 13) were the third most common set of change points in the bootstraps, occurring in 6.3\% of replicas, behind (2, 4, 10) ($12.9\%$) and (4, 5, 10) ($6.4\%$) (\cref{fig:fit}A). 
Fitting the model to each of $1000$ block bootstrapped resamples using individual faculty as the unit of resampling provides uncertainty estimates for all of the model's parameters. 
The relative instability of the inferred change point at year 13 is largely due to the fact that longer careers are less common in the data (full trajectories comprise only $510$ ($24.4\%$) of total trajectories, see Supporting Information); and only in the resamples with more of the full trajectories would the later career ages be detected as a change point. 
As a robustness check, we also fitted the full model to only the full trajectories, and find that the change point sets (4, 7, 11) and (4, 7, 13) are much more common across bootstrap replicas ($23\%$ in total). 

Within the maximum likelihood career stages estimated from the full model (4, 7, 13), the estimated variances in the pre-tenure early career $\widehat{\alpha_1} = 4.5$ (95\% CI: $[4.3, 4.6]$), $\widehat{\alpha_2} = 4.3$ (95\% CI: $[4.1, 4.4]$) were higher than the variances in the later career $\widehat{\alpha_3} = 3.8$ (95\% CI: $[3.7, 3.9]$), $\widehat{\alpha_4} = 3.5$ (95\% CI: $[3.4, 3.7]$). 
Meanwhile, the estimated $\beta_i$ parameter, which determines the mode of the career-stage Laplace distribution in conjunction with the productivity, fluctuated in an uncorrelated way with the average productivity, as did the mode $\mu_i$ when we fit it directly as a robustness check.
This finding confirms the insights from the simplified model: the fitted full model produces the canonical trajectory through changes in variance, rather than changes in the typical productivity. 
Hence, counter-intuitively, the distribution of the number of papers that a researcher is likely to produce in the next year (given their current year’s productivity) does not need to shift across a career in order to produce the aggregate pattern observed in the canonical trajectory. 
Rather, the canonical pattern can emerge merely from mid-career reductions in the variance in annual productivity.

\textbf{Canonical trajectory.} 
If the fitted full model includes the most salient aspects of individual productivity dynamics, then we expect simulations from the model to be statistically similar to the empirical trajectories. 

First, we examine whether the model simulations display a canonical trajectory in aggregate. 
Indeed, our simulated trajectories evolve similarly to empirical productivity trajectories on average, successfully recovering the rapid rise and gradual decline~(\cref{fig:fit}A). 
In fact, the average productivity is closely aligned between simulated and empirical trajectories, such that the largest average within-year difference between the two is less than one unit of productivity across an entire faculty career. 
This level of agreement is particularly notable because the model was fitted to individual level data, and yet it produces synthetic time series that yield the same aggregate pattern as the empirical data.

\begin{figure*}[htbp!] 
\centering 
\includegraphics[width=0.9\linewidth]{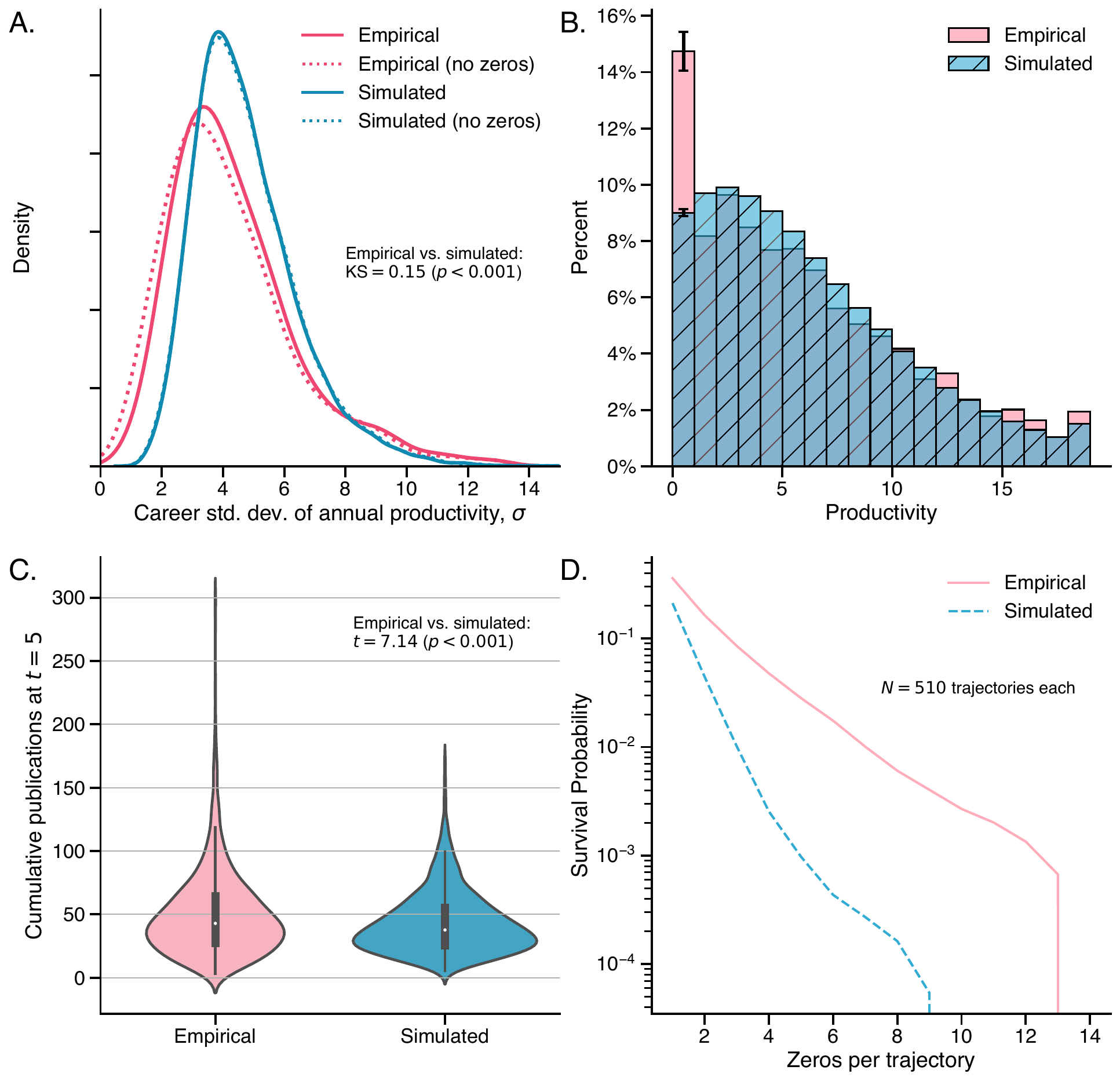} 
\caption{{\bf Comparing the random walk model to empirical data.} 
(A) Distributions of within-career standard deviations of productivity, for full empirical and simulated trajectories showing that empirical productivity variation tends to be slightly smaller ($\text{KS}=0.14$; $p < 0.001$), even if we omit zeros. 
(B) The distribution of annual productivities (full trajectories), showing a close match for all values except at zero between empirical and simulated careers. 
Black bars indicate the binomial 95\% Wald confidence intervals for the probability of zero publications. 
(C) Distributions of productivity of empirical and simulated trajectories at career year 5. 
Inside the violin plot, the white circle indicates the median, the thick bar indicates the interquartile range, and the thin bar indicates the centered 95\% containment interval. 
By career year 5, the simulated trajectories tend to have fewer publications than the empirical trajectories on average ($t=0.16$; $p < 0.001$), and the difference is especially pronounced among the tail of the most productive individuals. 
(D) Distributions of career years with zero publications within full empirical and simulated trajectories. 
The distribution of simulated and empirical trajectories with exactly one zero is similar, but more empirical trajectories exhibit more than one zeros than the simulated trajectories. 
}%
\label{fig:unfit} 
\end{figure*} 

\textbf{Career year of greatest productivity.} 
The year of greatest productivity is not directly parameterized by the random walk model. 
To evaluate the model's accuracy on this pattern of productivity, when fitted to the full trajectories only, we examine the distribution of the year in which a trajectory reaches its maximum productivity for the full trajectories and for $10,000$ trajectories simulated from the fitted model. 
We find that these two distributions~(\cref{fig:fit}B) are statistically indistinguishable~($KS=0.04, p=0.44$), indicating that the model naturally explains this pattern in the data. 

\textbf{Variance within and across careers.} 
Focusing on the full trajectories and computing the variance and standard deviations of productivity within each empirical and simulated trajectory, we find that the empirical trajectories tend to exhibit slightly lower variance than simulated trajectories~($KS=0.21, p<0.001$, \cref{fig:unfit}A). 
The prevalence of years with zero publications in empirical trajectories, however, is not sufficient to explain this difference~(\cref{fig:unfit}A). 

Empirically, faculty produce more cumulative papers by career year 5 than do simulated trajectories~($t=9.16, p < 0.001$, \cref{fig:unfit}C). 
This discrepancy is driven by a longer tail of cumulatively productive individuals in the empirical data who are not reproduced by the model: since researchers' productivity is lower variance than our model predicts~(\cref{fig:unfit}A), researchers with higher productivity are more consistently highly productive as well.

\textbf{Years with zero publications.} 
Comparing the empirical and simulated productivity distributions of the full trajectories, we observe that years with zero publications are substantially more common in the empirical data~(15\% vs 9\%, \cref{fig:unfit}B). 
Across empirical and simulated trajectories, the proportion of careers with exactly zero or one year of zero publications is similar, but empirical trajectories tend to have more zeros per trajectory than simulated ones~(\cref{fig:unfit}D). 
We note that the prevalence of years of zeros cannot be explained due to data quality issues within DBLP (see Supporting Information), and hence this discrepancy suggests that the dynamics that occur around a non-publishing state are not currently captured in our random walk model. 

\textbf{\rr{Cumulative productivity.}}
\rr{An observation and prediction made by William Shockley in 1957 is that cumulative productivity of scientific careers follows a lognormal distribution~\cite{shockley1957statistics}. Indeed, we find agreement within our model simulations that cumulative productivity at the end of a 21-year career are lognormally distributed~(\cref{fig:shockley}A), while the empirical data has a longer left tail~(\cref{fig:shockley}B; see Supporting Information for details).
In contrast to Shockley, however, our model operates on different assumptions, so our result offers an alternative explanation for the observation of cumulative lognormal productivity that does not rely on individual differences (see Discussion).}

\subsection{\rr{Validation across fields}}

\begin{figure}[hpb]
    \centering
    \includegraphics[width=0.8\linewidth]{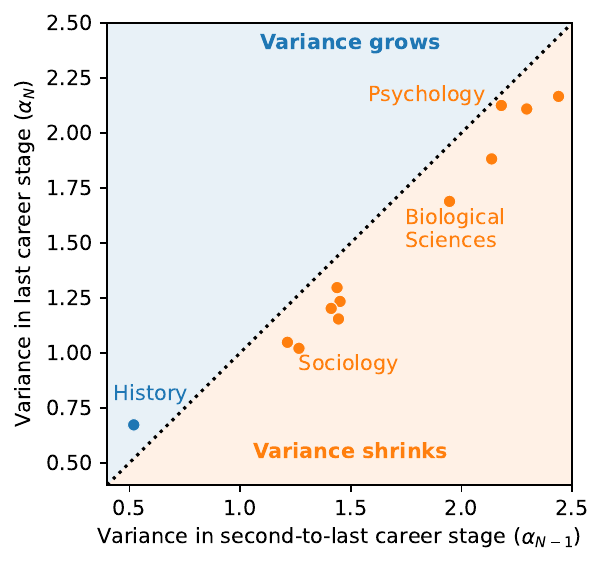}
    \caption{\rr{\textbf{Change in variance across fields} For each of the 12 fields shown in \cref{fig:aarc_trajectories}, we plot the fitted Laplace distribution variance of the productivity increments in the second-to-last inferred career stage (x-axis) with the last career stage (y-axis). The diagonal indicates a field where variance remained constant between the two career stages. Fields where variance shrank between the second-to-last and last career stage are plotted in orange below the
            diagonal, and fields where variance grew are plotted in blue above the diagonal. Several fields are labeled for illustration.
    }}
    \label{fig:variance_across_fields}
\end{figure}

\rr{Although the canonical trajectory has been observed across a variety of contexts in the literature, our specific model of decreasing variance across career stages has not been firmly established outside of Computer Science (but see~\cite{hamermesh2023older}).
    Moreover, the specific mechanism remains difficult to untangle when looking at a single field, since the average reduction of productivity after year six could be due to a field-level explanation or tenure.
To assess whether the inferred career stages roughly align with tenure times in the US, and that changes of variance can explain post-tenure reductions in productivity, we fit the same model to 12 different fields using data from the Academic Analytics Research Center (AARC).
Our fitted trajectories are at most 11 years due to the need to infer faculty start years from panel data (see Data).}

\rr{Across 12 different fields, the canonical trajectory appears broadly but not uniformly~(\cref{fig:aarc_trajectories}).
For every one of these fields except Management, our model infers a changepoint within one year of career age 7 (zero-indexed 6), which is the typical year after faculty undergo tenure review, consistent with a tenure-based explanation for changes in variance (see Supporting Information).
Consistent with our explanation that changes of variance can explain declining productivity after the early career, we find that the variance shrinks between the second-to-last career stage and the last career stage in all but one of the fields: history~(\cref{fig:variance_across_fields}).
These results indicate broad evidence for our hypothesis that a random walk with a change of variance can explain productivity trajectories across fields.
}

\section{Discussion} 

Scientific understanding about large-scale patterns in faculty productivity has been overly focused \rr{on} average phenomena, such as the canonical trajectory, rather than on the dramatic variability of individuals. 
This focus has drawn the field to incomplete theories of scientific productivity, such as individual-level theories that posit an increase and decline of individual capabilities (e.g., scientific creativity and energy) over the course of a career, that attempt to explain the canonical average without accounting for the environmental determinants of productivity~\cite{way2019productivity, zhang2022labor} or the broad diversity of real scientific trajectories. 
This empirical diversity of real productivity patterns poses a major challenge to all individual-level theories of scientific productivity, because it requires a successful theory to explain both the average pattern as well as the large variations across and within individuals. 

In this work, we discover two previously unknown statistical regularities: one in the distribution of early-career productivity and one in the distribution of year-to-year fluctuations in productivity~(\cref{fig:criteria}). 
We leverage these regularities to create a parsimonious explanation of productivity as a random walk where the variance in step size itself changes in a specific way across career stages. 
The model recapitulates both the canonical trajectory in average productivity and many empirical characteristics of the diversity of individual trajectories.
\rr{One surprising phenomena that our model reproduces is William Shockley's observation that cumulative research productivity within an institution tends to follow a lognormal distribution~\cite{shockley1957statistics}.
Shockley posited that such a distribution arose from variation in the attributes of individual scientists, but our model shows that such assumptions about individual variability are not necessary to reproduce the cumulative lognormal distribution.
Instead, identical individuals with randomly evolving productivity can produce cumulative lognormal productivities.}
These results, as well as the career statistics that the model does not fully reproduce, constitute a new perspective of scientific productivity and faculty careers rooted in randomness. 

The key insight of this model---that a random walk with high variance in the early career followed by decreased variance in the later career can produce the canonical trajectory in aggregate while maintaining high individual diversity---highlights a critical open question: what drives this decrease in variance from the early to the later career of \rr{academic faculty}? 
A sociological explanation for the higher early career variance focuses on the structure of faculty career incentives: acquiring research grants, forming research groups, and publishing papers constitutes a critical component of tenure evaluation, so faculty are pressured in their early career to accelerate their research output in a short timespan, in a way unlike in the later career when the ``start up'' effects of an early career are more distant~\cite{cole1979age}\rr{, a dynamic
also present in countries outside of the US and Canada~\cite{checchi2021incentives}}.
\rr{The literature on innovation suggests that the longer time horizons and stable job security of academic tenure may encourage exploration and risk-taking~\cite{manso2011motivating, azoulay2011incentives}, and evidence shows that productivity declines after tenure~\cite{brogaard2018economists, tripodi2025tenure}.}

For senior researchers, having an existing research group makes it more difficult to expand as much in relative terms---e.g., to quadruple the number of active group researchers from four to sixteen is much more challenging than to grow from one to four.
Established researchers can also be more selective about grant applications to avoid the logistical difficulties of managing a rapidly expanding and contracting group\rr{, and stable publication rates can occur out of habit~\cite{hamermesh2023older}.}
Additionally, in the later career, faculty have access to many more career paths than do early-career faculty, such as major university service roles related to curricular design and university administration, and scholarly service like editorships and professional society leadership, while the requirements for receiving tenure force all junior researchers into a narrower set of paths~\cite{cole1979age}.

The existence of research groups and career roles point toward latent structure that is more complex than our model. 
Random walks are Markovian, or ``memoryless'', in that this year's productivity only depends on the prior year's productivity. 
In addition, faculty who enter research inactive career roles can be expected to exhibit more years with zero papers than what our simulation predicts, which is precisely what we find in the data~(\cref{fig:unfit}D). 
By contrast, graduate student, postdoctoral, and research staff contracts are generally longer than a year~\cite{zhang2022labor}, meaning that a researcher's group size constitutes an unobserved latent variable that decreases the variance in faculty productivity. 
Both research groups and research inactive career roles reduce the variance in faculty productivity relative to a random walk, and indeed we observe slightly lower variances within empirical careers than what our model predicts~(\cref{fig:unfit}A), and higher cumulative variances across faculty~(\cref{fig:unfit}C). 
Even if individual productivity is more correlated across time than a memoryless model predicts, the discrepancy due to research inactive states is practically small relative to the remaining variance within careers~(\cref{fig:unfit}A).
Nevertheless, future work could model latent variables such as research group size and faculty research roles directly using a hidden Markov Model to potentially capture these non-Markovian aspects of productivity trajectories~\cite{zhang2022labor}.

The relationship between faculty retention and productivity is complex, and may potentially filter the data that we observe, especially in the full trajectory data.
Faculty leave tenure-track positions for a variety of reasons, such as workplace climate, work-life balance, and professional reasons, and these reasons interact in complicated ways with attrition~\cite{spoon2023retention}.
Attrition can happen among highly productive scholars who are pulled into industry positions, less productive scholars who fail to secure tenure, or average scholars who leave for non-professional reasons.
For the results that we can compute using all of the trajectories, the corresponding analyses using the full trajectory data produce qualitatively similar outcomes, suggesting that the role of attrition on the findings are negligible.

The dynamical approach we construct here effectively subsumes more specific mechanistic models, and poses a further puzzle for researchers: why does faculty productivity follow such clear mathematical distributions (the exponential distribution for early-career productivity, the Laplace distribution for year-to-year changes in productivity\rr{, and the lognormal distribution of cumulative productivity}), and why does a simple random walk model reproduce so many features of the empirical data, 
despite ignoring the main heterogeneities in academic careers such as prestige~\cite{wapman2022quantifying, way2019productivity}, gender~\cite{lariviere2013bibliometrics, li2022untangling, laberge2024genderedhiring}, parenthood~\cite{morgan2021unequal}, race~\cite{hoppe2019topic}, socioeconomic status~\cite{morgan2022socioeconomic}, and subfield~\cite{laberge2022subfield}?

One answer is that those heterogeneities are a subset of a panoply of contingent factors---tasks fundamental to the production of science such as delays in funding, student recruiting, peer review, coordination with collaborators including students, and regular variation due to the nature of research itself (experiments, data collection, computation, mistakes, dead ends, etc), not to mention non-academic sources of randomness, such as unexpected or 
variable life events---which are so numerous and unpredictable that together they constitute the bulk of the variation in productivity over time, giving rise to the appearance of dominating randomness.
\rr{In other words, randomness is a shorthand for explaining variance that would be implausible to fully model through data.}
Indeed, the Laplace distribution can appear when heterogeneous random walks are themselves aggregated together~\cite{kotz2001laplace}. 

The close agreement between the empirical data on changes in annual productivity and a Laplace distribution, which is symmetric, highlights a striking fact: the probability that a scientist's productivity increases next year by some amount very nearly equals the probability that it also decreases by the same amount in the following year. 
An interesting direction of future work would be to untangle the underlying factors and contingencies that make the distribution so symmetric.
Ultimately, any symmetry between increases and decreases in productivity is imperfect, because scientists cannot produce fewer than zero papers any given year. 
This ``hard'' boundary plays a crucial role in explaining how an increase in productivity variance becomes an increase in average productivity. 
That is, when annual productivity is close to zero, the zero boundary censors the distribution of changes in productivity, and that censoring shifts the average displacement upward~\cite{feller1991introduction}.
The higher the distribution's variance, the greater the censoring effect, and the larger the induced upward shift in the average change.
In this way, the zero boundary induces a coupling between the variance in the distribution of changes to productivity with the average productivity itself. 

\rr{Two potential policy implications flow directly from our findings.
First, researchers who become research inactive tend to remain so.
While our results do not differentiate between researchers who choose to become research inactive and those who do not, the possibility that some researchers become stuck in inactivity simply through chance suggests potential policy interventions for recovering potential research contributions from seasoned researchers.
In particular, grants or other mechanisms to assist faculty in research inactive career roles to transition back to research could prevent their subsequent contributions from becoming permanently lost to science.
Second, a small proportion of researchers may become highly productive simply through randomness and early-career luck, without any necessary differences in individual aptitude.
Scientific awards and other prestigious institutions designed to motivate scientific achievement should be thoughtful about the extent to which they are rewarding prior success.
Our work therefore adds to a growing body of literature on the importance of the Matthew Effect in understanding scientific career trajectories~\cite{perc2014matthew,bol2018matthew, petersen2011matthew}.}

\rr{The quantitative study of scientific careers has been approached by many scholars, typically using techniques from a social science methodological toolkit that aim to identify averages behind a veil of variance.}
Our results, based on a \rr{dynamical} model that centers this variability, show that changes in variance drive changes in the average, and that incentives and other system-level factors constrain and shape the way the fluctuations at the local level generate the aggregate trends.
\rr{Our work suggests a shift in perspective: individual-level fluctuations are an inherent part of research productivity, and the panoply of contingent factors are an inherent part of the system to be understood rather than averaged away.}
This shift toward randomness and \rr{variance}, away from deterministic laws, illuminates the broad diversity that characterizes real productivity patterns, within and across scientific careers.

\section{Acknowledgments}
\rr{We thank the Academic Analytics Research Center (AARC) for providing data.}
Funding: This work was supported in part by an Air Force Office of Scientific Research Award FA9550-19-1-0329 (NL, DBL, AC), an NSF Graduate Research Fellowship Award DGE 2040434 (SZ), and the NSF Alan T. Waterman Award SMA-2226343 (DBL). 
The complete data and source code for replicating the analysis and figures are available on Github (\href{https://github.com/samzhang111/faculty-trajectories}{https://github.com/samzhang111/faculty-trajectories}) 

\section{Author contributions}
SZ: Conceptualization, data curation, formal analysis, investigation, methodology, software, validation, visualization, writing; 
NL: Data curation, writing; 
SFW: Data curation;
DBL: Conceptualization, data curation, funding acquisition, resources, visualization, writing; 
AC: Conceptualization, data curation, funding acquisition, investigation, methodology, project administration, resources, supervision, visualization, writing.

\bibliographystyle{unsrt}
\bibliography{trajectories}

\clearpage

\setcounter{figure}{0}
\renewcommand{\thefigure}{S\arabic{figure}}

\setcounter{table}{0}
\renewcommand{\thetable}{S\arabic{table}}

\setcounter{equation}{0}
\renewcommand{\theequation}{S\arabic{equation}}

\begin{widetext}
\begin{center}
{\Large\bfseries Supplementary Material}\\[1em]
{\large Scientific Productivity as a Random Walk}\\[1em]
Sam Zhang, Nicholas LaBerge, Samuel F. Way, Daniel B. Larremore, Aaron Clauset
\end{center}
\end{widetext}

\vspace{2em}

\section{\rr{Analysis on other fields}}
\begin{figure*}[htpb]
    \centering
    \includegraphics[width=1\linewidth]{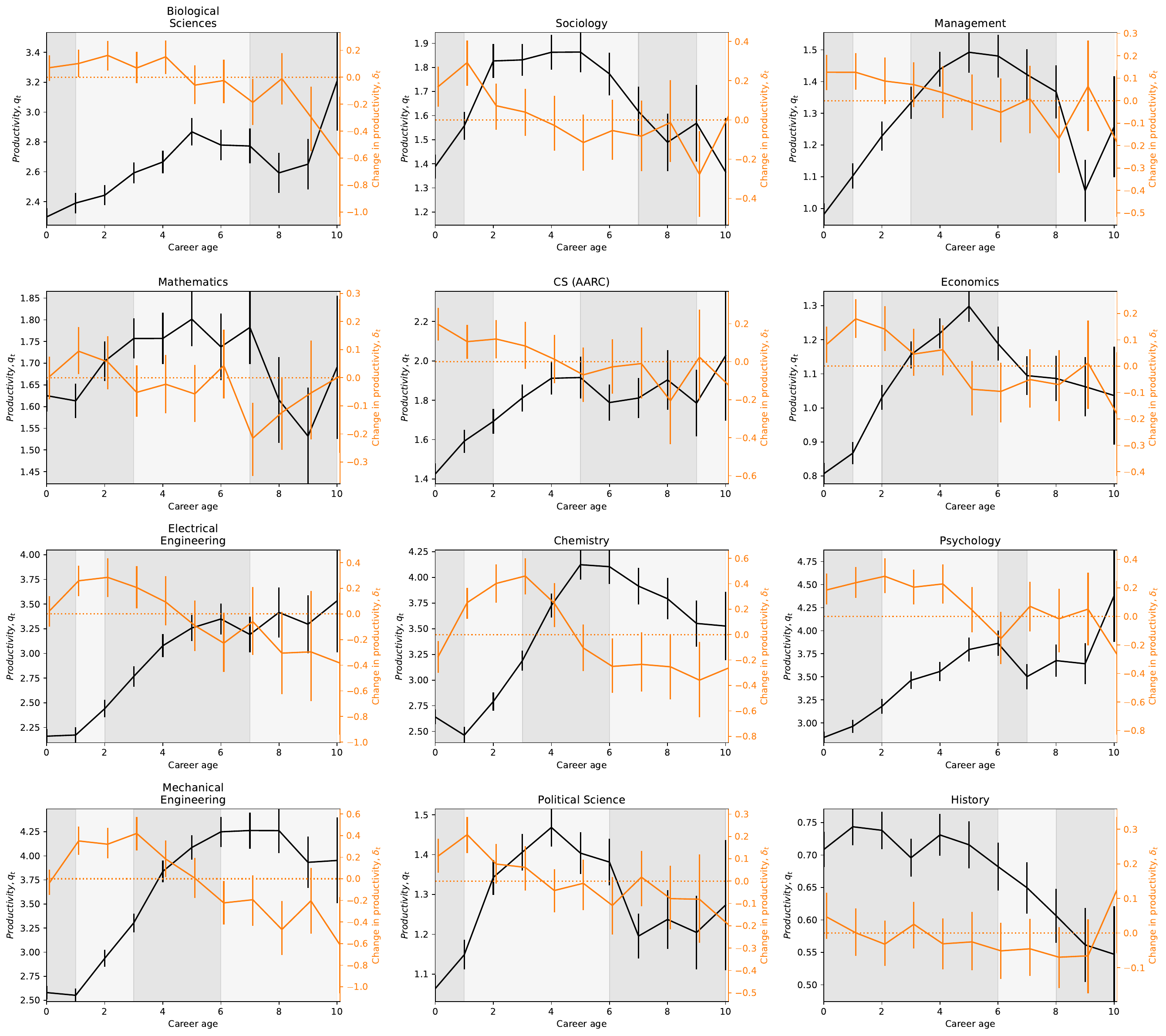}
    \caption{\rr{\textbf{Productivity trajectories across fields} For each of twelve fields, we show in black (left axis) the average productivity by career age, and in orange (right axis) the average change in productivity. A dashed orange line indicates the baseline of no change in productivity. Error bars indicate 1 SE. Alternating dark and light gray background indicates the inferred career change stages.
    }}
    \label{fig:aarc_trajectories}
\end{figure*}

\rr{To test whether our model transcends disciplinary boundaries or is limited to computer science, we additionally analyze data across 12 other fields using data provided by the Academic Analytics Research Center (AARC), which includes comprehensive faculty employment records between 2011 and 2023, with publications linked in Scopus.
The academic fields that we include in the AARC analysis are biological sciences, sociology, management, mathematics, economics, electrical engineering, chemistry, psychology, mechanical engineering, political science, and history. We further include computer science as a validation check for our existing analysis, and we excluded physics after an initial examination due to outliers from high-energy physics.
Unlike the complete hand-collected census of Computer Science tenure-track faculty in the US and Canada that we employed in the main manuscript, the AARC data does not contain the year that faculty start their academic jobs. Instead, we rely on the relative completeness of the AARC dataset to infer faculty start years based on the year that they first appear in a given department. After excluding individuals without any records linked in Scopus, we result in 102,473 years of data across 22,952 individuals, ranging from 1,204 scholars in sociology to 2,669 in psychology.
The canonical trajectory appears broadly but not uniformly across fields~(\cref{fig:aarc_trajectories}), confirming prior observations using AARC and Scopus to study productivity trajectories~\cite{tripodi2025tenure}.}

\rr{Our trajectories differ from Ref.~\cite{tripodi2025tenure} in that we align researchers at the start of their careers (``career age''), due to our interest in the evolution of trajectories starting from the beginning of a faculty position, as opposed to indexing trajectories based on the year that faculty transition from assistant to associate professors (``tenure age''). If productivity trajectories are better explained by career ages or tenure ages, then the population of all faculty should exhibit lower variance within any given career or tenure age.
    To the contrary, we find that the within-year variance of productivity is roughly the same on average when indexing either by career age or tenure age~(\cref{fig:tenure}). This is consistent with a pluralistic approach of understanding productivity trajectories: both the starts of careers and promotion times are critical moments in a researcher’s career that structure productivity. These results point toward the difficulty of disentangling the respective effects of career age and tenure. Further work
could examine the mechanics of tenure by focusing on midcareer movers who do not achieve tenure at their first institution (although this would be a biased sample of individuals who may not reflect the overall research population), or as the reviewer suggests, comparing high-quality faculty career datasets from other countries with different promotion systems.}

\section{\rr{Relative importance of tenure}}
\begin{figure}[htpb]
    \centering
    \includegraphics[width=0.8\linewidth]{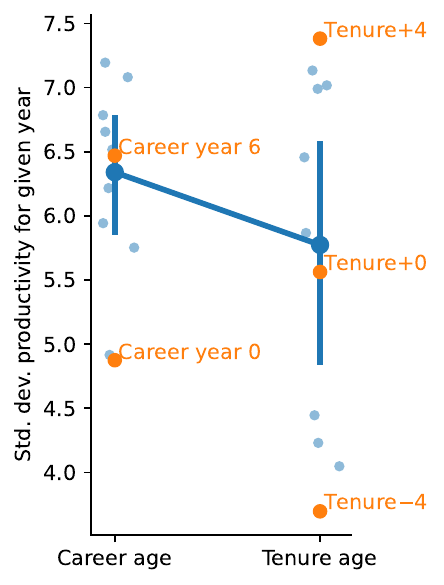}
    \caption{\rr{\textbf{Variance explained by indexing with career age vs. tenure age} Standard deviations of productivity across all faculty indexed by year since appointment in department (career age) vs. faculty indexed by year that their title transitions from assistant to associate professor (tenure age). Thick blue circles are averages across years, with 95\% CIs. Orange dots are arbitrary years labeled as examples.
    }}
    \label{fig:tenure}
\end{figure}

\rr{Fundamentally, the difficulty of estimating the relative effect of tenure is that within the US and Canada, assistant professors who remain employed for much longer than seven years at a single department without promotion are far and away the exception, due to academic norms that expect either promotion or termination.
From observational data, we lack a comparison set of individuals who remain at an institution at the rank of assistant professor.
Exceptions to that pattern may be too singular or confounded to assist us in drawing broader causal conclusions, such as individuals who experience extended medical leave, elite universities where tenure occurs at promotion to full professor, and midcareer transitions where faculty restart as assistant professor at a second institution.}

\rr{In line with that typical alignment between career age and tenure age, our analysis did not find early evidence that tenure is more explanatory for productivity trajectories than simply their career ages.
For instance, if tenure alone explained the variance-shift mechanism, we should hope that indexing faculty careers by time-to-tenure would create year-indices with less variance of productivity than indexing by career age (i.e., that time-to-tenure “explains” productivity more than time-since-career-start). That expectation did not hold among the population of faculty who we observed both starting and being promoted between 2011–2023 in AARC---there was no statistically significant difference
(\cref{fig:tenure}).}

\rr{The most plausible explanation is not that career age and tenure age are fully synchronized: faculty do undergo tenure review earlier or later than the mode for various reasons. Rather, both career age and tenure age are good explanations for productivity trajectories for different reasons, and indexing on either event produces ranges where variance of productivity is reduced (in the early career for career age, where the sharp increase in average productivity year-on-year is such a clear
dynamic that in 10 of the 12 fields we analyzed, a changepoint was detected at the end of the first or second year (Fig. R3); and post-tenure for the tenure age, where the evidence is convincing that faculty reduce their output, likely because they focus on more novel and impactful work~\cite{tripodi2025tenure}.
A potential extension to this model could therefore incorporate promotion as a latent variable that occurs around year 6, which we would expect to further capture some of the variance associated with that change point. However, pushing the model to simulate a full counterfactual where a faculty member never receives tenure but remains at an institution as an assistant professor is too far outside of the support of the empirical data.}

\section{Modeling details}

\subsection{The Model} 

We model a researcher's annual productivity as a stochastic process with the following assumptions: 

\medskip 
\textbf{Assumption 1:} The number of publications $q_t$ in a given year $t$ cannot be negative. 

\textbf{Assumption 2:} Initial researcher productivity $q_0$ follows an exponential distribution~(\cref{fig:criteria}A). 

\textbf{Assumption 3:} The change in productivity from year to year $\delta_t = q_{t+1} - q_t$ is distributed independently of previous years' productivity, and follows a Laplace distribution~(\cref{fig:criteria}B-D). 

\textbf{Assumption 4:} Researcher productivity exhibits different dynamics in different career stages~(\cref{fig:criteria}E). 
\medskip 

Assumption 3 implies that the stochastic process is Markovian.
These assumptions do not uniquely define a specific model.
For example, the boundary condition around the combinations of Assumptions 1 and 3 could be plausibly modeled with either a reflecting or truncated boundary, and we proceed with a truncated boundary for computational simplicity.
We present two possible models that fulfill these assumptions: first, a simplified model, and then, a full model. 

\textbf{Simplified model.} We fix the mode of the Laplace distribution globally to an arbitrary constant ($\mu=-1$) and fit the first-year exponential rate parameter using the empirical mean of the data, $\widehat{\lambda}_0=4.65$. 
Then we separate the data into two career stages: years $0$ to $4$, representing the ``early career'' stage, and years $5$ to $20$, representing the remainder of a career. 
Each career stage has the same location $\mu=-1$, but different scale parameters $\alpha_1$ and $\alpha_2$, respectively, which are the only free parameters of this simplified model. 
We then simulate trajectories from the simplified model,
and assess whether individual simulated trajectories exhibits the canonical pattern using a simple model selection approach across a family of linear and piece-wise linear regressions.
We systematically explore the variance parameter space of this model, and then plot the fraction of trajectories that meet the criteria for being canonical to produce~\cref{fig:simplified_canonical}.

We follow Ref.~\cite{way2017misleading} in using model selection to classify individual trajectories as canonical.
    We fit a two-part piecewise linear model to each trajectory for each possible change point between career years 3 and 17, as well as a linear model.
    We then use the small-sample correction of the AIC (AICc) to determine the best-fitting model.
    The trajectory is labeled as canonical if the best-fitting model is a piecewise model (that is, the linear model is not the best fit) that additionally fulfills the following criteria: the slope of the first piece is positive, the slope of the second piece is negative, and the magnitude of the slope of the first piece is at least twice the magnitude of the slope of the second piece.

\textbf{Full model.} Rather than assuming only two career stages at a fixed change point, in the full model we allow up to four change points, whose locations are estimated from the data. 
And we allow both the location and scale parameters of the Laplace distribution to vary with each change point. 
Lastly, instead of using a constant mode, for mild technical reasons, we estimate a slope parameter $\widehat{\beta}$ for each career stage, where $\widehat{\mu} = \widehat{\beta} q_t$, which allows for sharper changes within career stages, which can now be arbitrarily short~(\cref{fig:qt_vs_delta}). 

A career stage $i$ is defined in terms of change points $c_i, c_{i+1}$, where we construct the data within each career stage $D_i = \lbrace (q_t, q_{t+1}) \rbrace$, where $c_i \le t < c_{i+1}$. 
Each candidate set of career stages is identified by a tuple of change points $(c_1, c_2, c_3)$, where $c_3$ or both $c_2$ and $c_3$ may be omitted. We implicitly assume a change point at the two endpoints for career age $0$ and at $t_{\text{max}}$, and we adopt the convention that the career stages include the right endpoints. 
For instance, the change point set $(2, 5)$ would encode career stages 0--2, 3--5, and 6--20. 
This procedure yields 1,159 possible sets of one, two, or three change points from 1 to 19. 

For each career stage $i$, we estimate the model's parameters using an alternating optimization. 
First, we estimate a global mode $\widehat{\mu}_g$ of the truncated Laplace distribution by identifying the mode of $\mathbb{P}(\delta_t)$. 
Second, since the log-likelihood of the Laplace is then both smooth and convex in the scale parameter for each career stage $\alpha_i$, we perform maximum likelihood estimation on $\alpha_i$ using $\widehat{\mu}_g$ as the mode. 
Third, we parameterize the log-likelihood in terms of a slope $\widehat{\beta}$, rather than location, for each career stage, writing $\mu_i(q_t) = \beta_i q_t$, and setting $\alpha_i = \widehat{\alpha}_i$ from the previous step. 
Finally, we re-estimate $\alpha_i$ using $\widehat{\mu}_i = \widehat{\beta}_i q_t$.

We estimate separate parameters for each career stage. 
A career stage $i$ is defined in terms of change points $c_i, c_{i+1}$, where we construct the data within each career stage $D_i = \lbrace (q_t, q_{t+1}) \rbrace$, where $c_i \le t < c_{i+1}$. 
In order to accommodate the flexibility of empirical career structures while remaining computationally feasible, we consider every possible set of four or fewer career stages, or equivalently, one, two, or three change points. 
We select career stages via model selection using the AIC to correct for overparameterization. 
Thus each candidate set of career stages is identified by a tuple of change points $(c_1, c_2, c_3)$. 
For each career stage $i$, we estimate the model's parameters using an alternating optimization. 

We estimate a single exponential parameter \mbox{$\widehat{\lambda}_0 > 0$} for the distribution of $q_0$, which is shared across all models and is independent of the choice of change points. 
To generate a synthetic productivity trajectory, we first generate a choice of initial productivity $q_0$ from the estimated model $\mathbb{P}(q_0 | \lambda_0)$. 
Then we set $q_{t+1} = q_t + \delta_t$ for $c_i < t \le c_{i+1}$ which we draw from the parameterized distribution $\mathbb{P}(\delta_t | q_t, \alpha_i, \beta_i)$. 
The synthetic productivity trajectories simulated from the model are used to perform the comparisons with the empirical data shown in \cref{fig:fit} and \cref{fig:unfit}.

\subsection{Validity of the Markovian assumption}

The Markov assumption implies that 

\begin{align}
    P(q_{t+1} = r_{t+1} &| q_t = r_t, q_{t-1} = r_{t - 1})\\
    =P(q_{t+1} = r_{t+1} &| q_t = r_t).
\end{align}

\begin{figure}[htb!] 
\centering 
\includegraphics[width=1\linewidth]{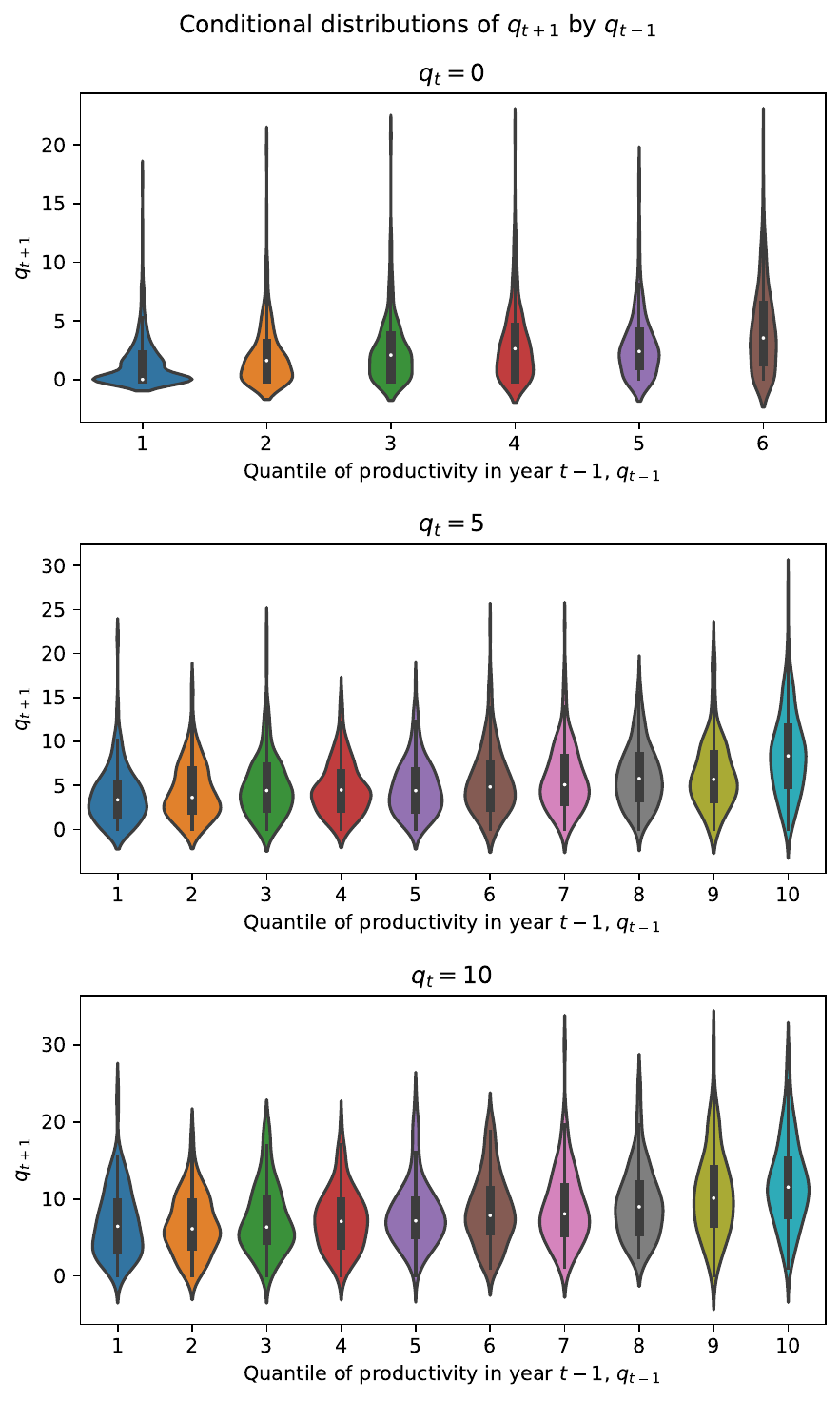} 
\caption{\textbf{Conditional distributions of productivity.} 
    Violin plots showing the conditional distribution and box-and-whiskers of quantiles of productivity, given the previous year productivity.
    Conditional distributions are shown as kernel density estimates, and the boxes display quartiles of the data, with the median as a white dot.
    The whiskers are the farthest datapoint within 1.5 times the interquartile range from the nearest hinge.
    Higher quantiles are more productive.
} 
\label{fig:conditional_distributions} 
\end{figure}

By conditioning on the intermediate value $q_t$, we can observe whether $\mathbb{P}(q_{t+1})$ varies by the value in the previous timestep, $q_{t-1}$~(\cref{fig:conditional_distributions}).
We find confirmation for two of the results stated in the main paper.
First, compared to simulated trajectories, the empirical trajectories have less variance within careers, greater variance across careers, and many more years with zeros~(\cref{fig:unfit}).
Here we confirm this observation from the point of view of the Markov property: the overall trend of increasing means implies that historically high (resp. low) values lead future values to be high (resp. low), even after conditioning on intermediate timesteps~(\cref{fig:conditional_distributions}).
In the main paper, we find significantly greater abundance of years with zero papers empirically than our model predicted, even after correcting for missingness in the data~(\cref{fig:unfit}B,D).
We still see that the non-Markovian behavior is strongest around zero (that is, conditioning on $q_t=0$, we see that trajectories where $q_{t-1}=0$ are far more likely to produce zeros in year t+1 than other values of $q_{t-1}$), but nevertheless there is a smaller non-Markovian trend in the other conditional distributions as well, particularly in the top-most quantiles.
This provides validation for our explanation for how empirical trajectories reach higher cumulative productivities by year 5 than the simulations would predict~(\cref{fig:unfit}C).

\subsection{The distribution of the random walk increments}

\begin{figure*}[htpb]
    \centering
    \includegraphics[width=1\linewidth]{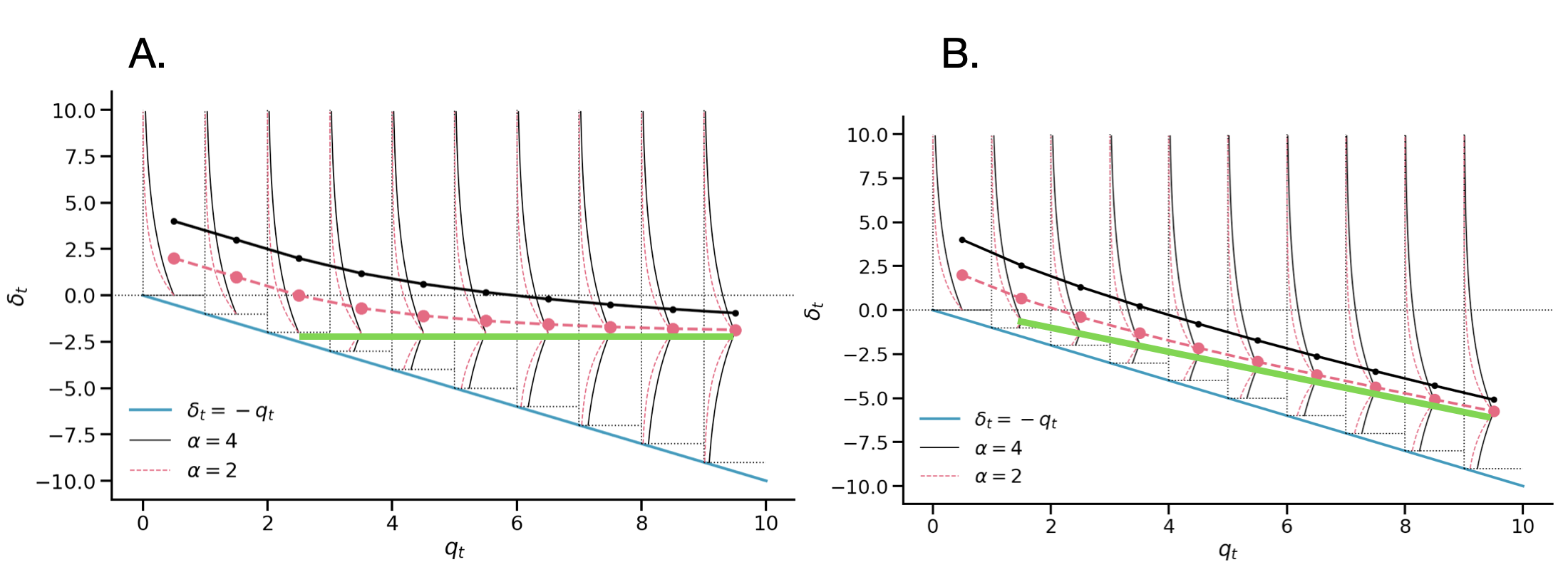}
    \caption{\textbf{Distribution of increments given productivity.}
        (A) For two Laplace distributions with the same estimated mode ($\mu=-2$), the means (horizontal lines) of the distribution with higher variance ($\alpha=4$, solid black) are higher than the means of the distribution with lower variance ($\alpha=2$, dashed pink).
        (B) A similar diagram, except the location is parameterized in terms of $\beta$, where $\mu=\beta q_t$. Here $\beta=-1$. The difference in location is emphasized on the plots as thick green lines, where in (A) the line is constant at $\mu=-2$, while in (B) the line has a negative slope of $\beta=-1$.
}
    \label{fig:qt_vs_delta}
\end{figure*}

For an observation $( x, \delta ) := ( q_t, q_{t+1} - q_t )$, the conditional pdf from the Laplace distribution with a fixed mode $\mu$~(\cref{fig:qt_vs_delta}A) is:

\begin{equation*}
    f( x, \delta ) =  \mathbb{I}\lbrace \delta > -x \rbrace ( 2 - e^{ - \mu/\alpha })^{-1} ( 1/\alpha ) e^{-|\delta - \mu|/\alpha}
\end{equation*}

On the other hand, the conditional pdf from the Laplace distribution with a productivity-dependent mode $\beta$ where $\mu = \beta x$~(\cref{fig:qt_vs_delta}B) is given by:

\begin{equation*}
f( x, \delta ) =  \mathbb{I}\lbrace \delta > -x \rbrace ( 2 - e^{ - \beta x/\alpha })^{-1} ( 1/\alpha ) e^{-|\delta - \beta x|/\alpha}
\end{equation*}

We employ the fixed mode $\mu$ for the simplified model, since it is simpler, and the productivity-dependent mode $\beta$ for the full model, since it produces a more qualitatively successful fit of the empirical data.
In particular, using the $\mu$ parameterization, the tails of the annual fluctuations are sufficiently heavy that the convergence rate is unrealistically slow for the timescales representing real productivity trajectories, and the $\beta$ parameterization allows for faster convergence.

The estimated $\beta_i$ parameters from the full model are less than $1$ in every career stage, even the first stage where average productivity increased annually, meaning that typical annual productivity decreases compared to the prior year's productivity. 
In other words, faculty appear to experience a general ``drag" in maintaining their productivity, which we interpret as asymmetry between decreasing productivity (requiring only inaction) compared to increasing productivity (requiring an increase in resources and effort). 
It is higher variance, made asymmetric by the hard boundary at productivity zero, that overcomes this drag and effectively prevents faculty productivity from slumping toward zero. 

\section{Recovering known parameters from synthetic data} 
\label{sec:synthetic}

\begin{figure}[htpb]
    \centering
    \includegraphics[width=1\linewidth]{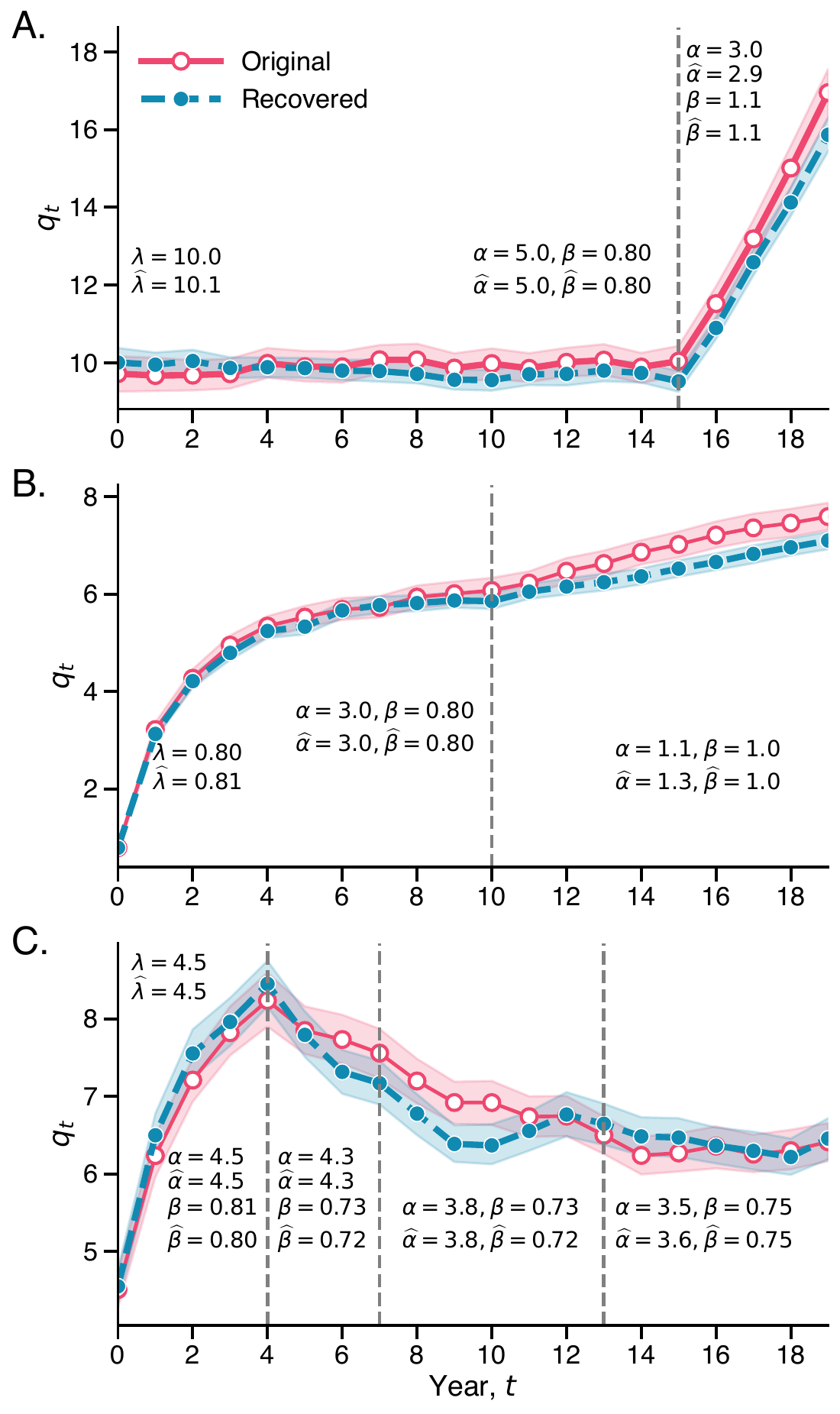}
    \caption{\textbf{Model recovery on synthetic data with known structure.}
        Parameter estimates and simulated productivity trajectories for three qualitatively different specifications of the generative model, where
        (A) stages are delineated by sharp and linear changes, 
        (B) by less sharp but nonlinear changes, or
        (C) simulated trajectories that were generated using the parameters estimated from the empirical data, illustrating that in realistic parameter regimes we can reasonably recover the same parameters.
}
    \label{fig:recovers_simulations}
\end{figure}

Using synthetic trajectories with known parameters, the full model correctly recovers the relevant parameters generally up to an error of $\pm0.1$ for $\lambda$ and $\alpha$, and $\pm0.01$ for $\beta$~(\cref{fig:recovers_simulations}).
To put that in perspective, an error in $\lambda$ of $0.1$ corresponds to a tenth of a publication in a scientist's first year, and the errors in $\alpha$ and $\beta$ can be interpreted on a similar scale.
This scale of error is practically negligible compared to the much larger underlying variability in productivity across individuals.

Trajectories generated from the fitted model closely align with the original synthetic data across a qualitatively diverse range of scenarios with a varying number of change points (Fig.~\ref{fig:recovers_simulations}).
    The inferred change points match the change points used to generate the original synthetic data, and we select the correct number of change points in each of these instances, both in situations with a sharp cutoff like in scenario (A), as well as when change points are more subtle as in scenarios (B) and (C).

\section{DBLP Data}

This random walk model can be applied to any dataset composed of time series of individual researcher productivities that range onward from the beginning of a career, e.g., the first year of a permanent research position like a tenure-track job at a PhD-granting institution. 
Datasets of individual-level productivity derived from bibliographic databases like the Web of Science~\cite{clarivate2022wos} are attractive because of their scale, but pose additional complexities due to the need to first disambiguate individuals, then stratify by fields, which can exhibit widely different average productivity levels~\cite{piro2013macro}, and finally stratify by different roles, e.g., faculty vs. trainees, or researchers employed at institutions with different research intensities. 

We avoid such complexities by studying a dataset of known computer science faculty at PhD-granting institutions in the US or Canada~\cite{way2017misleading}, which we linked to their publications as recorded in DBLP~\cite{dblp2016november}, a bibliographic database that focuses on computing and which was used to identify the underlying diversity of productivity trajectories~\cite{way2017misleading}. 
The faculty dataset was a complete collection of the 205 department or school-level academic units on the Computing Research Association’s Forsythe List of PhD-granting departments in computing-related disciplines in the US and Canada~\cite{clauset2015systematic, cra2012forsythe}.
The manual collection process yielded 5,032 faculty in these units, and the dataset was subsetted to the 2,583 whose PhD degree and first assistant professorship appointment were at units in the sample.
The DBLP was then joined to the faculty census, using manual name disambiguation as necessary, yielding 2,453 faculty with linked publication records.
The further inclusion criteria described in the ``Data'' section of the main manuscript resulted in the reported 2,085 faculty in our analysis.

Current faculty often begin publishing before the start of their faculty career, but the underlying dynamics of productivity are different in the pre-faculty stage of a career~\cite{zhang2022labor}. 
We focus our analysis here only on productivity during a faculty career, and so we exclude those publications from years prior to that career's beginning. 
Although DBLP has reasonably good coverage over most venues in which computer science faculty publish, it is not complete, and faculty working in subfields that are not well-indexed by DBLP, such as interdisciplinary computing, would appear as anomalously unproductive researchers. 

In order to correct for rising individual productivity and uneven historical coverage, Way et al. adjusted historical productivities using a linear model estimated from publication data manually extracted from CVs~\cite{way2017misleading}.
Such a linear adjustment preserves years with zero papers, which may lead to an over-estimate in the number of years with zero publications.
To evaluate this possibility, we randomly select eight individuals among those with at least ten years of zero publications, and manually counted the number of years when they have zero publications in both DBLP (108) and on their websites and Google Scholar (82).
This tally yields a maximum likelihood binomial estimate that $75.9\%$ (95\% CI: $(67.9\%, 84.0\%)$) of DBLP-observed zero-publication years are correct.
With the most conservative correction ($67.9\%$ of years with zero publications are truly zero), and most conservative estimate for the empirical mean number of zeros ($14.1\%$ at the bottom of the estimated 95\% CI), we would still find that $9.5\%$ of years in the full empirical trajectories are years with zero publications.
This estimate excludes the 95\% CI of the mean number of zeros in 10,000 simulated trajectories $(8.5\%, 8.7\%)$, implying that zero-productivity years are likely more common in real productivity trajectories than our model can account for.

\section{\rr{Skewness and kurtosis of productivity increments}}
\rr{To further diagnose differences between the empirical DBLP data and the simulated full model output, we computed the skewness and kurtosis within each of the three inferred career stages for both empirical and simulated trajectories, constructing 95\% block bootstrap CIs at the person level for the DBLP data, and regular 95\% bootstrap CIs for the simulated trajectories. We found skewness to be consistently higher in simulated than empirical trajectories. In the first career stage from year 1 to 4, the
skewness was $0.232$ (95\% CI [$0.133$, $0.327$]) for empirical data and $0.765$ (95\% CI [$0.696$, $0.834$]) for simulated data;
from years 5--7, the skewness was $0.141$ (95\% CI [$0.024$, $0.246$]) empirically and $0.689$ (95\% CI [$0.606$, $0.776$]) for simulation;
and from years 8--end, the skewness was $0.034$ (95\% CI [$-0.053$, $0.123$]) empirically and $0.721$ (95\% CI [$0.679$, $0.760$]) for simulation.
By contrast, kurtosis did not exhibit a consistent pattern: it was greater in the model than in the data during the first two stages, but the reverse held in the final stage. Specifically, in years 1--4, empirical kurtosis was $2.292$ (95\% CI [$1.827$, $2.662$]) compared to $2.998$ (95\% CI [$2.670$, $3.332$]) in the model; in years 5--7, it was $2.360$ (95\% CI [$1.735$, $2.834$]) empirically versus $2.941$ (95\% CI [$2.512$, $3.507$]) in the model; and in years 8--end, it was $4.024$ (95\% CI [$3.302$, $4.599$]) empirically compared to $2.929$ (95\% CI [$2.724$, $3.143$]) in the model.}

\section{\rr{Connection to Shockley model of productivity}}
\rr{Our model results in the surprising recapitulation of a pattern on research productivity introduced by William Shockley in 1957~\cite{shockley1957statistics}.
    Shockley found that over the course of a variety of scientific careers, the cumulative productivity within institutions tended toward a lognormal distribution, and had theorized some possible mechanisms, namely, that scientific papers are the result of combining individual traits or skills in such a way that the success of the project depends on each of these traits or skills.
    In other words, these traits or skills are connected in series, so that the outcome depends multiplicatively on each of these components.
    Then, if these traits or skills were each normally distributed, their product would be lognormally distributed.}

\begin{figure*}[htpb]
    \centering
    \includegraphics[width=1\linewidth]{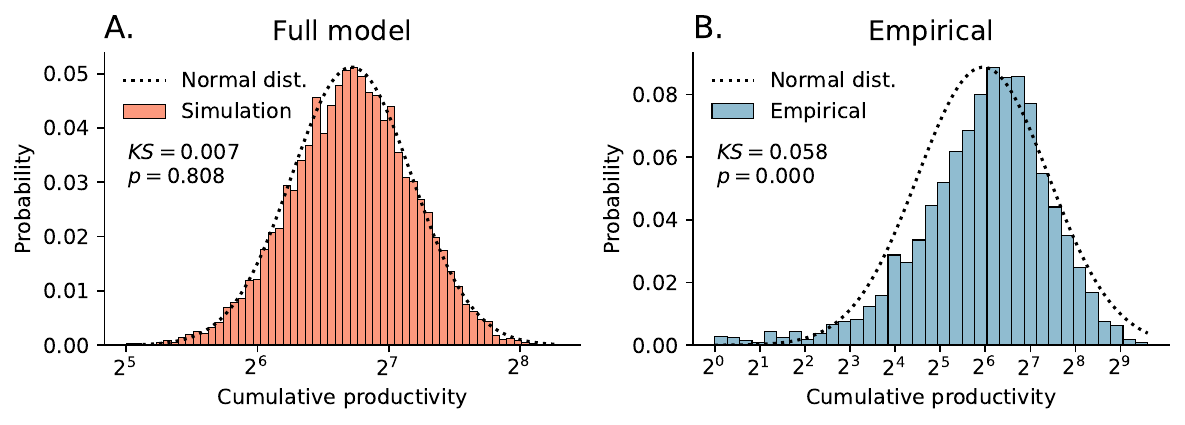}
    \caption{\rr{\textbf{Cumulative productivity distributions compared to fitted normal distributions.}
            (A) Cumulative productivities for 10,000 simulated careers of 21 years from full random walk models compared with fitted normal distribution, and 2-sided KS test statistic with p-value, using the dataset of all DBLP trajectories. (B) Cumulative productivities and fitted normal distributions of all empirical trajectories.
        }}
    \label{fig:shockley}
\end{figure*}

\begin{figure*}[htpb]
    \centering
    \includegraphics[width=1\linewidth]{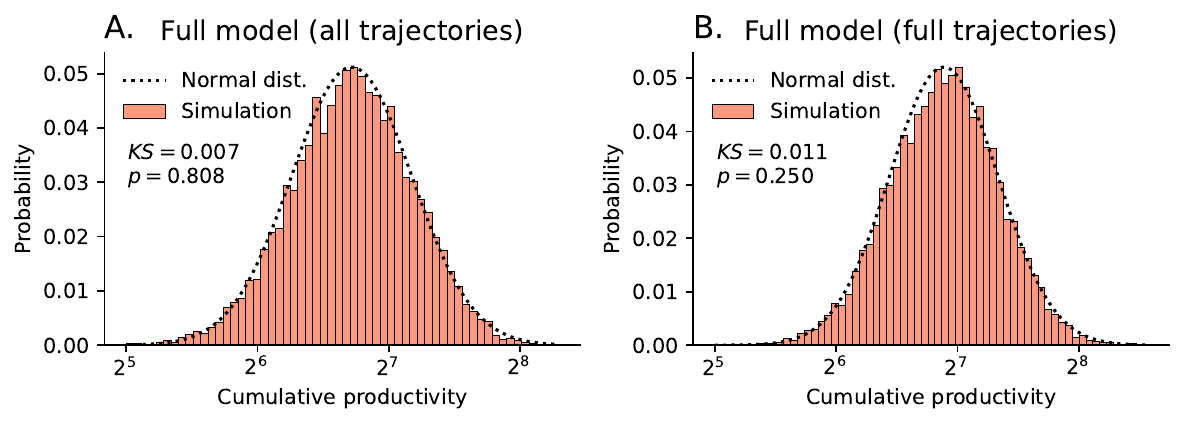}
    \caption{\rr{\textbf{Cumulative productivity distributions compared to fitted normal distributions.}
            (A) Cumulative productivities for simulated careers from full random walk models compared with fitted normal distribution, and 2-sided KS test statistic with p-value, using the dataset of all DBLP trajectories. 
            (B) Cumulative productivities for simulated careers from full random walk models compared with fitted normal distribution, and 2-sided KS test statistic with p-value, only using the dataset of full DBLP trajectories. 
        }}
    \label{fig:shockley_full}
\end{figure*}

\rr{Interestingly, productivity trajectories simulated from our full model also produce cumulative productivities over a 21-year span that are statistically indistinguishable from a lognormal distribution (\cref{fig:shockley}A). By contrast, the empirical data is not immediately lognormal, due to a surfeit of less productive careers (\cref{fig:shockley}B). The same pattern holds when using simulations from the model only fitted to full 21-year-long careers (the full data;
    \cref{fig:shockley_full}). Shockley also found a wider tail than lognormal empirically, however, which he overcame through manually selecting a cut-off through ``trial
    and error''.}

\rr{Our model poses potential theoretical challenges to Shockley’s model. Shockley anticipated and argued against the ``organizational hypothesis'' that coauthorship produces the lognormal distributions of cumulative productivity. However, he did not consider the possibility that identically skilled individuals can achieve cumulative productivities that are lognormally distributed simply through the variability accrued by careers progressing through multiple career stages. Indeed, our model presents
a potential alternative to Shockley’s preferred explanation of differences in individual aptitude.}

\section{\rr{Baseline models}}
\subsection{\rr{Exponential distribution of papers around time-varying mean}}
\rr{Here we motivate our random walk approach model by directly mathematizing a model posited in the “canonical trajectory” literature: the ``aging functions'' introduced by Ref.~\cite{bayer1977career}.
In particular, we model a time-varying mean with exponential distribution of papers around that mean, with independent productivity across timesteps conditioned on the mean.
We note that the difference between two iid exponential distributions is a Laplace distribution, providing the model with some initial credibility.}

\rr{We model to changing mean following Ref.~\cite{way2017misleading}, by assuming that the average trajectory is either linear or piecewise-linear with two components.
    We choose between the linear model or the specific piecewise-linear model using model selection.
    In particular, we fit a piecewise linear model with two pieces on every possible changepoint between years 3 and 17 for the entire dataset of full 21-year trajectories, as well as a linear model without any changepoint,
and select the model out of that family with the best AICc.}

\begin{figure*}[htpb]
    \centering
    \includegraphics[width=1\linewidth]{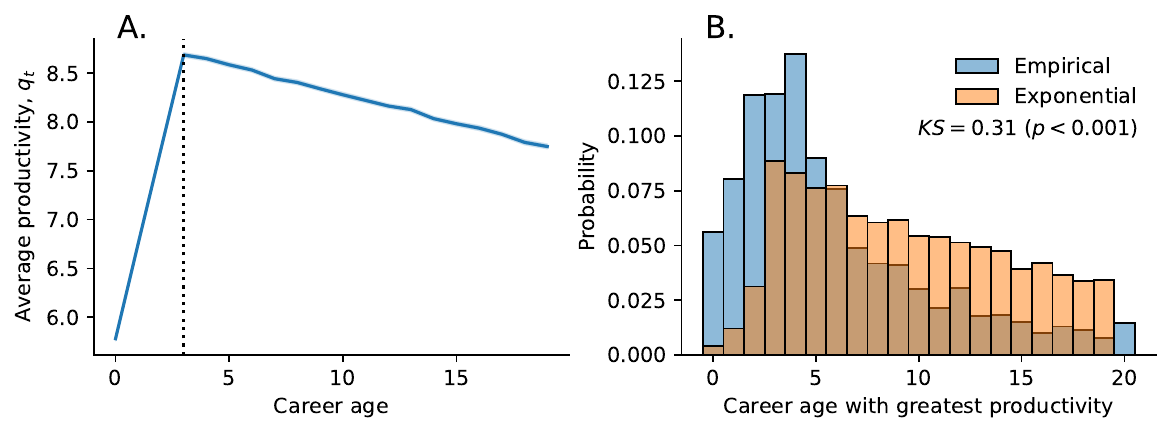}
    \caption{\rr{\textbf{Minimal model with exponentially-distributed productivity around a time-varying mean.}
            Minimal model with exponentially-distributed productivity around a time-varying mean. (A) Average career productivity for each career age, simulated from the fitted model. The vertical dotted line indicates the inferred changepoint, and error bands indicate 95\% CIs. (B) Distributions of years with greatest productivity within each (blue) empirical and (orange) simulated trajectory. Histograms are normalized such that each bar represents the probability that a given trajectory has its year of greatest productivity within that bin.
        }}
    \label{fig:minimal_model1}
\end{figure*}

\rr{The best fitting model displayed a changepoint at the fourth career year, with initial intercept at $4.77$ papers and pre-year-4 slope of $0.98$ more papers/year (\cref{fig:minimal_model1}A). After year 4, the slope changes to $-0.06$. As expected, this fulfills the criteria from Ref.~\cite{way2017misleading} of being a canonical trajectory, since the first slope is positive, the second slope is negative, and the absolute value of the first slope is at least twice that of the absolute value of the second slope.
We ignore the fact that the simulated average trajectory does not look precisely like the empirical data, since through semi-parametric approaches, one could theoretically fit the average trajectory as closely as one desires.
More concerning, however, and something that a semi-parametric approach would likely not resolve, is the disparity between the distributions of the most productivity years within the empirical data and in the simulation.
The simulated data from this exponential model is much more likely to present its most productive year later in the career than the empirical data, and respectively underestimate the probability of an early-career peak, despite the fact that its inferred changepoint was so early in the career ($KS=0.30$, $p < 0.001$; \cref{fig:minimal_model1}B).}

\subsection{\rr{Poisson-distributed maturation of synchronized projects}}

\begin{figure*}[htpb]
    \centering
    \includegraphics[width=1\linewidth]{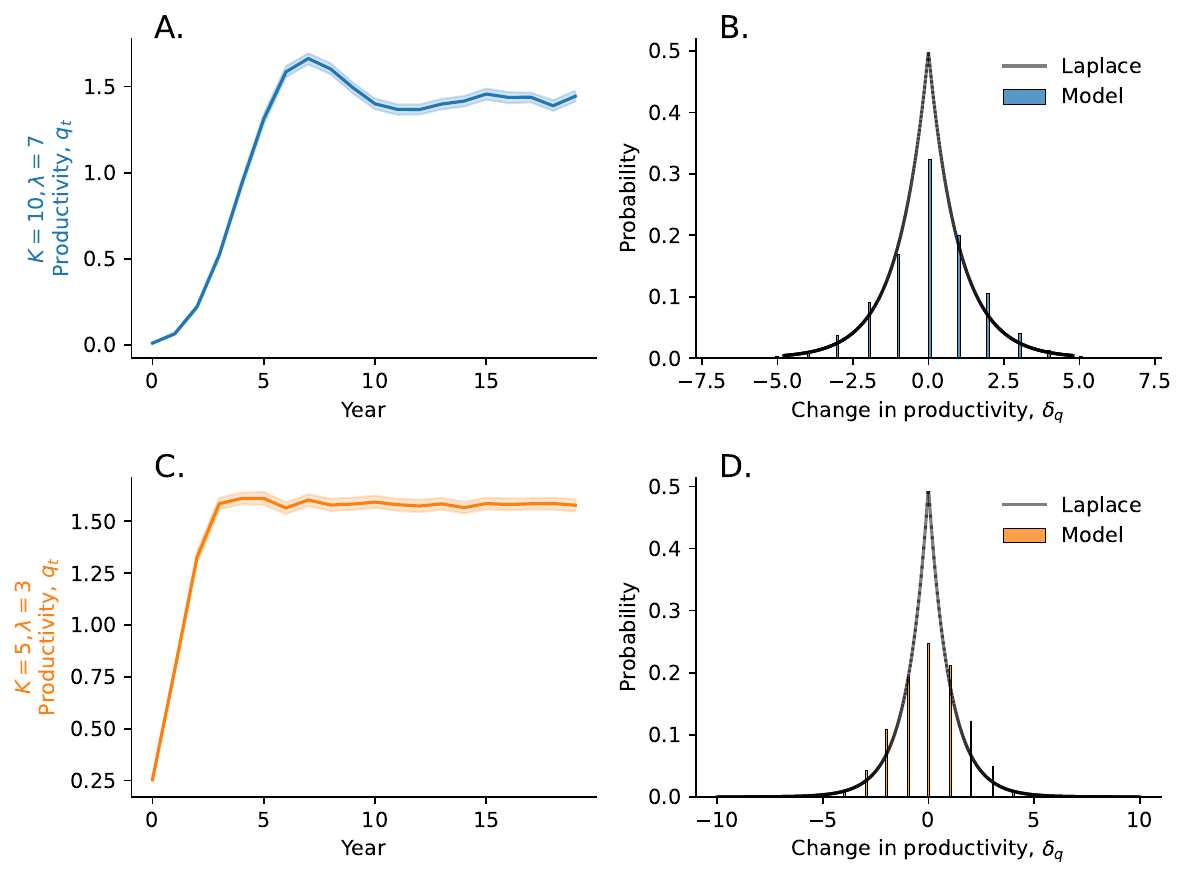}
    \caption{\rr{\textbf{Second simplified model: Poisson-distributed maturation of synchronized projects.}
            (A) Average productivity per year for simulated trajectories with $K = 10$, $\lambda = 7$. (B) Distribution of change in productivity increments compared with the fitted Laplace distribution for the (10, 7) model. (C) Average productivity per year for simulated trajectories with $K = 5$, $\lambda = 3$. (D) Distribution of change in productivity increments compared with the fitted Laplace distribution for the (5, 3) model.
        }}
    \label{fig:minimal_model2}
\end{figure*}

\rr{In a second simplified model, new faculty initiate $K$ projects at the same time, each of which finishes (yielding a paper) according to a Poisson distribution with rate $\lambda$. Once a project is finished, it resets.
Simply through the synchronization and subsequent decoherence of the initial basket of projects, one can observe trajectories that appear almost canonical, for example, if faculty work on 10 projects each of which mature with rate 7~(\cref{fig:minimal_model2}A).}

\rr{However, in more sensible ranges of the parameter space, where a researcher embarks upon 5 projects and completes them on an average of 3 years, then the resulting trajectory is no longer canonical (\cref{fig:minimal_model2}C). Another discrepancy between this and empirical reality is that the productivity increments are far from symmetrically exponentially distributed (\cref{fig:minimal_model2}B,D). Despite the unsuitability of this second simplified model for modeling scientific productivity, we note
it as an interesting and natural way that canonical trajectories may emerge in other contexts.}

\section{\rr{Relevance for prediction of individual trajectories}}

\rr{Predicting individual productivity would certainly serve as an impressive validation of any model, but our work outlines the challenge faced by any such predictor: simply by structuring the year-to-year variability of faculty productivity on top of a very basic random walk model, there is enough richness to reproduce the canonical trajectory, the diversity of individual trajectories, and the Shockley cumulative productivity distribution. In other words, many interesting phenomena about
scientific productivity trajectories can be more parsimoniously explained by aspects of the variance, rather than the conditional expectations.
Indeed, Ref.~\cite{rorstad2015publication} performed a regression analysis of 12,400 Norwegian researchers’ productivity, controlling for gender, age, and position, and found that they could only explain 13.5 to 19\% of the variance in productivity.
Thus even with these crucial factors (and country, implicitly) controlled for, 80\% of individual productivity remains unmodeled.}

\rr{A focus on prediction collapses the variance into an average in two senses: first, the objective becomes to model the conditional expectation of productivity given covariates such as the prior productivity; second, the overall loss function, such as the root mean squared error, is typically an average of losses across individuals. Here our model is fundamentally more interested in predicting the overall distribution of productivity trajectories, including their
    outliers, rather than any individual trajectories alone. From the perspective of ``predicting the population'', we already include many predictive model checks, since we generate simulated trajectories, and statistically compare those trajectories with the empirical ones.
    For instance, we find it interesting that our model successfully predicts the occurrence of outliers that were not explicitly included in the model, such as the distribution of faculty whose most productive year is
very early or very late in their careers~(\cref{fig:unfit}B).}

\rr{From the point of view of conditional expectations, our model expects regression to the mean within the structure of the truncated Laplace distribution, such that within DBLP, no researcher has an expected productivity (given their prior year’s productivity) greater than 23 papers per year. Yet within our dataset, over 772 researcher-years exhibit productivity greater than 23 papers, including 26 highly productive researcher-years with over 50 papers.}

\rr{Our predictive model checks indicate that there is more inertia to faculty careers than our model would predict (Fig. 4A). For that reason, we may expect a simple baseline predictor where productivity is predicted to be exactly the same as the previous year to even outperform our model. Indeed, using 1000 bootstrap samples, we find that the average RMSE of the extrapolatory baseline across bootstraps is 837.96 (95\% bootstrap CI: [827.19, 848.37]), while predictions from our model yield an
average RMSE of 1054.09 (95\% bootstrap CI: [1035.66, 1071.96]). These predictive checks remind us that our approach to capturing the overall distribution by only modeling global temporal structure currently comes at the expense of certain individual variance that can be retained.}

\rr{When used for prediction, our model collapses the variance across individuals more than these simple baselines. If we apply a simple baseline of predicting a researcher’s output to be a constant average of their first few years of productivity (or their last observed productivity in the early career) then we have surrendered all within-individual variation, but preserved across-individual diversity. Our model, when deterministically applied using conditional expectations, performs a fixed
point iteration~(\cref{fig:qt_vs_delta}), thereby driving all faculty toward a constant productivity.}

\end{document}